\documentclass[referee]{aa} 
\input{psfig.sty}

\begin{document}

\title{Emission and absorption lines of gamma-ray bursts affected by
the relativistic motion of fireball ejecta}

\author{Yi-Ping Qin \inst{}}

\offprints{Yi-Ping Qin}

\institute{National Astronomical Observatories/Yunnan Observatory,
Chinese Academy of Sciences, P. O. Box 110, Kunming, Yunnan,
650011, P. R. China; ypqin@public.km.yn.cn}

\date{Received 29 November 2002 / Accepted 12 May 2003}

\titlerunning{Emission and absorption lines of gamma-ray bursts}

\authorrunning{Y.-P. Qin}

\maketitle

\begin{abstract}
We display by numerical calculation how rest frame spectral lines
appear in the observed spectrum of gamma-ray bursts due to the
Doppler effect in the fireball framework. The analysis shows that:
a) in the spectrum of a relativistically expanding fireball, all
rest frame lines would shift to higher energy bands and would be
significantly smoothed; b) rest frame weak narrow emission lines
as well as narrow absorption lines and absorption line forests
would be smoothed and would hardly be detectable; c) the features
of rest frame broad emission lines as well as both strong and weak
broad absorption lines would remain almost unchanged and therefore
would be easier to detect; d) deep gaps caused by rest frame broad
absorption lines would be significantly filled; e) a rest frame
emission line forest would form a single broad line feature; f)
the observed relative width of the rest frame very narrow line
would approach $ 0.162$; g) when the Lorentz factor $\Gamma $ is
large enough, the observed line frequency $\nu _{line}$ and the
rest frame line frequency $\nu _{0,line} $ would be related by
$\nu _{line}\approx 2\Gamma \nu _{0,line}$. We also investigate
the effect of time dependence of the line intensity and the effect
of variation of $\Gamma $. We find that the feature of rest frame
dimming narrow emission lines would disappear when $\Gamma $ is
very large. The form of emission lines would be sharp on both
edges when $\Gamma $ varies with time. This phenomenon depends not
only on the initial Lorentz factor but also on the observation
time. \keywords{gamma-rays: bursts
--- gamma-rays: theory --- radiation mechanisms: nonthermal ---
relativity}
\end{abstract}

\section{Introduction}

Due to the observed great output rate of radiation of gamma-ray
bursts (GRBs), most models of the objects envision an expanding
fireball (see e.g., Goodman 1986; Paczynski 1986). The gamma-ray
emission would arise after the fireball becomes optically thin, in
shocks produced when the ejecta collide with an external medium or
in shocks occurred within a relativistic internal wind (Rees \&
Meszaros 1992, 1994; Meszaros \& Rees 1993, 1994; Katz 1994;
Paczynski \& Xu 1994; Sari et al. 1996). As the expanding motion
of the outer shell of the fireball would be relativistic, the
Doppler effect must be at work. As pointed out by Krolik \& Pier
(1991), relativistic bulk motion of the gamma-ray-emitting plasma
can account for some phenomena of bursts. However, in many cases,
the whole fireball surface should be considered, and accounting
for the effect the fireball surface itself would play a role
(Meszaros \& Rees 1998; Qin 2002, hereafter Paper I).

As revealed by the observation of absorption lines of afterglows
(Metzger et al. 1997), GRBs are confirmed to be events occurring
within the environment of stars. Based on the assumption that
there might be a strong magnitude field within the fireball and
that the expansion would be relativistic, it was believed that
synchrotron radiation would become a dominate mechanism (Liang et
al. 1983). Unfortunately, the spectra observed so far are so
different that none of the mechanisms proposed can account for
most of the spectral data of the objects (Band et al. 1993;
Schaefer et al. 1994; Preece et al. 1998, 2000).

Based on the assumption of the environment of stars, mechanisms of
formation of emission as well as absorption lines for GRBs have
been well established. Meszaros and Rees (1998) suggested that
gamma-ray burst outflows may entrain small blobs or filaments of
dense, highly ionized metal-rich material, which might be
accelerated by the flow to Lorentz factors in the range 10---100.
In the event of neutron-star collisions or black hole-neutron star
collisions, neutrino-antineutrino annihilation can produce
electron-positron pairs. Thus it would be natural if there exist
in the outer shell of fireballs some high energy emission lines
such as the $6.4keV$ line and the $511keV$ annihilation line
(Ramaty \& Meszaros 1981). As reviewed by Piran (1999), both
absorption and emission features have been reported by various
experiments prior to BATSE. Absorption lines in the $20-40keV$
range were observed by several experiments. GINGA discovered
several cases of lines with harmonic structure (Murakami et al.
1988; Fenimore et al. 1988). These lines were interpreted as
cyclotron lines (reflecting a magnetic field of $\approx
10^{12}Gauss$). Emission features near $400keV$ were claimed in
other bursts (Mazets et al. 1980). However, so far, BATSE has not
found any of the spectral features (absorption or emission lines)
reported by earlier satellites (Palmer et al. 1994; Band et al.
1996). Meszaros and Rees (1998) pointed out that within the
relativistic fireball model an observed broadened spectral line
might be a blue-shifted iron X-ray line. We suspect that, in
detecting line features, the Doppler effect associated with the
expanding fireball might play a role.

In the following we will investigate what one should expect in the
spectrum if there are emission as well as absorption lines in the
rest frame of fireballs.

\section{Effect on narrow emission lines}

We consider a fireball expanding with a definite value of Lorentz factor $%
\Gamma $ and study only the core content of the Doppler effect in
the fireball framework with the cosmological as well as other
effects being temporarily ignored. The rest frame radiation is
assumed to be constant and independent of direction.

Let the fireball be observed at time $t$ and at frequency $\nu $.
Then the expected spectrum would be (Paper I)
\begin{equation}
\nu f_\nu (t)=\frac{2\pi \widetilde{R}^2(t)\nu }{D^2\Gamma ^3}\int_0^{\pi /2}%
\frac{I_{0,\nu }(t_{0,\theta },\nu _{0,\theta })\cos \theta \sin \theta }{%
(1-\beta \cos \theta )^5}d\theta ,
\end{equation}
with
\begin{equation}
\widetilde{R}(t)=\beta [c(t-t_c)-D]+R_c,
\end{equation}
where $D$ is the distance of the fireball to the observer; $\theta
$ is the angle to the line of sight; $\nu _{0,\theta }$ is the
rest frame emission
frequency of the differential surface, of the fireball, with $\theta $; $%
t_{0,\theta }$ is the proper emission time of the differential
surface; $t_c$ is any coordinate time concerned; $R_c$ is the
radius of the fireball at time $t_c$; and $I_{0,\nu }(t_{0,\theta
},\nu _{0,\theta })$ is the rest frame intensity of the
differential surface. Frequencies $\nu _{0,\theta }$ and $\nu $
are related by the Doppler effect. The proper time $t_{0,\theta }$
and observation time $t$ can be well linked by considering the
travelling of light from the fireball to the observer (see Paper
I).

Assume that during some period the radiation of the fireball is
dominated by a certain mechanism and within this interval of time
the rest frame radiation intensity can be expressed as:
\begin{equation}
I_{0,\nu }(t_{0,\theta },\nu _{0,\theta })=I_0(t_{0,\theta
})g_{0,\nu }(\nu _{0,\theta }),
\end{equation}
where $g_{0,\nu }(\nu _{0,\theta })$ describes the dominant
radiation mechanism while $I_0(t_{0,\theta })$ represents the
development of the intensity magnitude. For constant radiations,
$I_0(t_{0,\theta })=I_0$. In this case, the spectrum would be
\begin{equation}
\nu f_\nu (t)=\frac{2\pi I_0\widetilde{R}^2(t)\nu }{D^2\Gamma
^3}\int_0^{\pi /2}g_{0,\nu }(\nu _{0,\theta })\frac{\cos \theta
\sin \theta }{(1-\beta \cos \theta )^5}d\theta ,
\end{equation}
here the emission is assumed to cover the whole energy range.

As mentioned above, there is no single mechanism proposed that can
well represent all the observed spectra of GRBs. In practice, an
empirical form called the GRB model (Band et al. 1993) was
frequently, and rather successfully, employed to fit most burst
spectra (Ford et al. 1995; Preece et al. 2000). The relative
intensity of the GRB model is (Band et al. 1993)
\begin{equation}
g_{0,\nu ,G}(\nu _{0,\theta })=\{
\begin{array}{c}
(\frac{\nu _{0,\theta }}{\nu _{0,p}})^{1+\alpha _{0,G}}\exp
[-(2+\alpha
_{0,G})\frac{\nu _{0,\theta }}{\nu _{0,p}}]\quad (\frac{\nu _{0,\theta }}{%
\nu _{0,p}}<\frac{\alpha _{0,G}-\beta _{0,G}}{2+\alpha _{0,G}}) \\
(\frac{\alpha _{0,G}-\beta _{0,G}}{2+\alpha _{0,G}})^{\alpha
_{0,G}-\beta
_{0,G}}\exp (\beta _{0,G}-\alpha _{0,G})(\frac{\nu _{0,\theta }}{\nu _{0,p}}%
)^{1+\beta _{0,G}}\quad (\frac{\nu _{0,\theta }}{\nu _{0,p}}\geq \frac{%
\alpha _{0,G}-\beta _{0,G}}{2+\alpha _{0,G}}),
\end{array}
\end{equation}
where $\alpha _{0,G}$\ and $\beta _{0,G}$\ are the lower and
higher indexes, respectively. As an illustration, we shall adopt
in the following the GRB form with the typical values of the
parameters: $\alpha _{0,G}=-1$ and $\beta _{0,G}=-2.25$ (Preece et
al. 1998, 2000).

To show the effect on both emission and absorption lines, we
consider the following rest frame radiation:
\begin{equation}
g_{0,\nu }(\nu _{0,\theta })=\max \{g_{0,\nu ,G}(\nu _{0,\theta
})+\sum_{i=1}^ne_{0,\nu ,i}(\nu _{0,\theta }),0\}
\end{equation}
where $e_{0,\nu ,i}(\nu _{0,\theta })$ describes the relative
intensity of lines:
\begin{equation}
e_{0,\nu ,i}(\nu _{0,\theta })=\frac{h_i}{\exp (2+\alpha _0)}\exp [-\frac{%
(\nu _{0,\theta }/\nu _{0,p}-\nu _{0,line,i}/\nu _{0,p})^2}{2\times 10^{n_i}}%
].
\end{equation}

Now we consider two narrow emission lines and take
\begin{equation}
\nu _{0,line,1}/\nu _{0,p}=0.01,\qquad h_1=1000,\qquad n_1=-8
\end{equation}
and
\begin{equation}
\nu _{0,line,2}/\nu _{0,p}=100,\qquad h_2=1,\qquad n_2=0.
\end{equation}

\begin{figure}[tbp]
\vbox to 3.0in{\rule{0pt}{3.0in}} \includegraphics{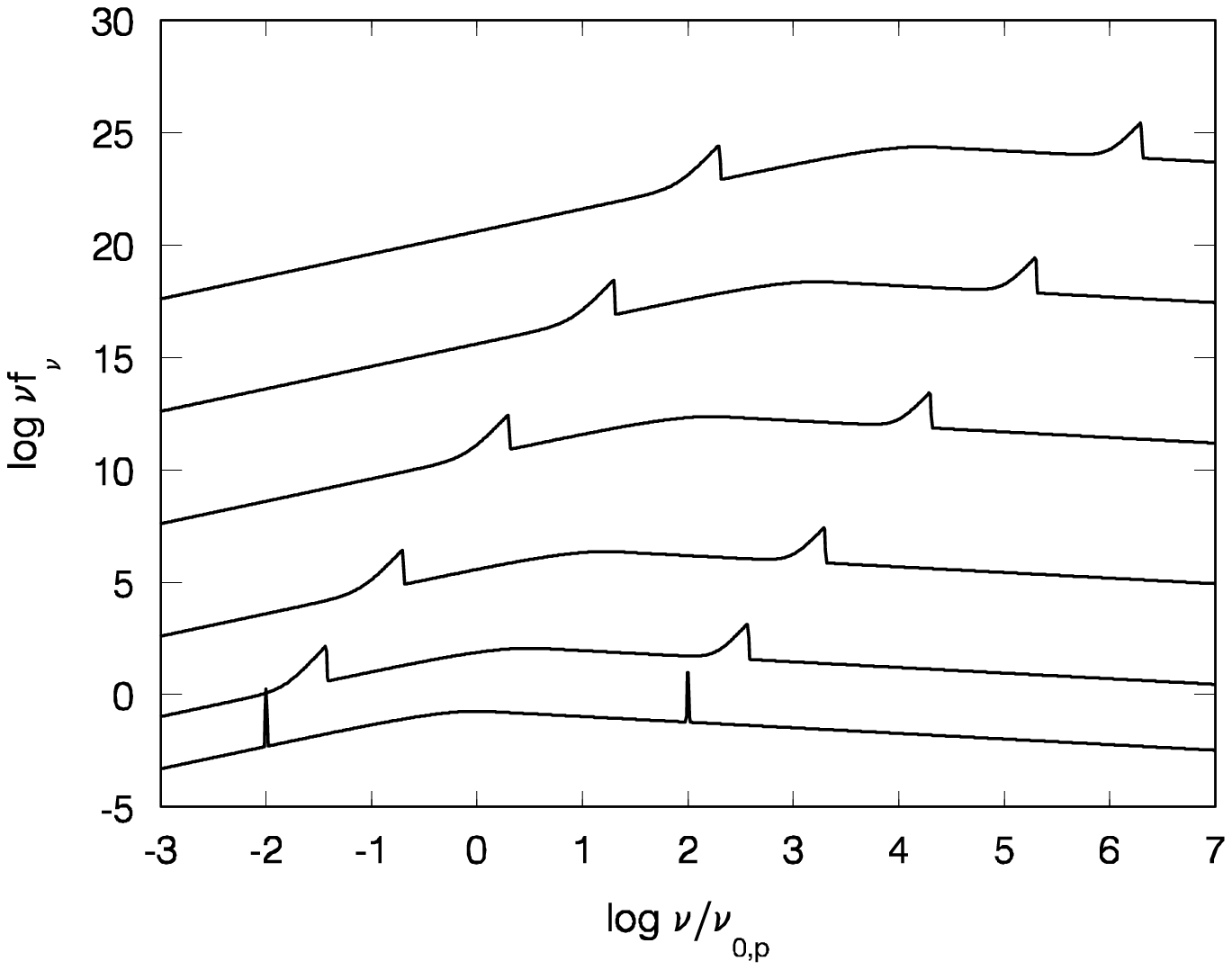} \caption{The
expected spectrum of a fireball with its rest frame radiation
being the GRB form of $\alpha _0=-1$ and $\beta _0=-2.25$ plus the
two
emission lines defined by (8) and (9), observed at time $t$, where we take $%
2\pi I_0\nu _{0,p}\widetilde{R}^2(t)/D^2=1$. The solid lines from
the bottom to the top correspond to $\Gamma
=1,2,10,100,1000,10000$, respectively.} \label{Fig1}
\end{figure}

Shown in Fig. 1 are the $\log (\nu f_\nu )-\log (\nu /\nu _{0,p})$
curves of the fireball containing in its rest frame radiation the
two emission lines, for various values of $\Gamma $. The figure
shows that, in the spectrum of a relativistically expanding
fireball, rest frame narrow emission lines indeed shift to higher
energy bands and are significantly broadened. Therefore, there
would be no narrow emission lines expected in the spectrum of the
object. In addition, a rest frame narrow emission line at high
energy bands would form an up-turning tip of the high energy tail
in the observed spectrum (an observed $MeV$ tip can come from a
rest frame $keV$ line when $\Gamma =1000$ or from a $10keV$ line
when $\Gamma =100$) (see Fig. 1).

To show the effect of the broadening, we calculate the relative
width of the frequency range related to the $FWHM$ of the line
feature, $\Delta \nu _{FWHM}/\nu _{line}$, where $\nu _{line}$ is
the observed line frequency at which the peak flux of the line
feature is found. Listed in Table 1 are the values of $\nu
_{line}$ as well as $\Delta \nu _{FWHM}/\nu _{line}$, coupled to
various values of $\Gamma $, for the two lines.

\begin{table}[ht]
\caption{List of $\nu _{line}$ and $\Delta \nu _{FWHM}/\nu
_{line}$ for the two narrow emission lines}
\begin{tabular}{lllll}
\hline\hline \multicolumn{1}{l}{} & \multicolumn{2}{l}{$\nu
_{0,line}/\nu _{0,p}=0.01$} & \multicolumn{2}{l}{$\nu
_{0,line}/\nu _{0,p}=100$} \\
\hline
$\Gamma $ & $\nu _{line}/\nu _{0,p}$ & $\Delta \nu _{FWHM}/\nu _{line}$ & $%
\nu _{line}/\nu _{0,p}$ & $\Delta \nu _{FWHM}/\nu _{line}$ \\
\hline $1\times 10^0$ & $1.00\times 10^{-2}$ & $2.35\times
10^{-2}$ & $1.00\times
10^2$ & $2.35\times 10^{-2}$ \\
$2\times 10^0$ & $3.65\times 10^{-2}$ & $1.79\times 10^{-1}$ &
$3.65\times
10^2$ & $1.79\times 10^{-1}$ \\
$1\times 10^1$ & $1.95\times 10^{-1}$ & $1.86\times 10^{-1}$ &
$1.95\times
10^3$ & $1.86\times 10^{-1}$ \\
$1\times 10^2$ & $1.96\times 10^0$ & $1.86\times 10^{-1}$ &
$1.96\times 10^4$
& $1.86\times 10^{-1}$ \\
$1\times 10^3$ & $1.96\times 10^1$ & $1.86\times 10^{-1}$ &
$1.96\times 10^5$
& $1.86\times 10^{-1}$ \\
$1\times 10^4$ & $1.96\times 10^2$ & $1.86\times 10^{-1}$ &
$1.96\times 10^6$ & $1.86\times 10^{-1}$ \\
\hline
\end{tabular}
\end{table}

We find that $\nu _{line}$ is a linear function of $\Gamma $,
approximately following $\nu _{line}\approx 2\Gamma \nu
_{0,line}$, when $\Gamma $ is large enough (say, $\Gamma
>10$). Therefore, if $\nu _{line}$ is identified, then one can
take it as an indicator of the expansion speed of the fireball. It
also shows that, in the case discussed above, where the relative
width of the rest frame narrow emission lines is $0.0235$ (see the
case of $\Gamma =1$ in Table 1), the relative width of the line
feature becomes almost one magnitude larger than that of the rest
frame emission line when the fireball expands relativistically.
When $\Gamma $ is large
enough, the relative width would approach an asymptotic value (here it is $%
0.186$).

As illustrated by an analytical calculation (see Appendix A), $\nu
_{line}$ of a very narrow rectangle emission, which is taken to
represent a very narrow emission line, can be determined by $\nu
_{line}=\nu _{0,line}/\Gamma (1-\beta )$. When $\Gamma \gg 1$, it
approaches $2\Gamma \nu _{0,line}$, which is the same as that
shown in Table 1. In this situation, the value of $ \Delta \nu
_{FWHM}/\nu _{line}$ can be estimated by $0.162+0.415\triangle \nu
_{0,line}/\nu _{0,line}$. When $\triangle \nu _{0,line}/\nu
_{0,line}\rightarrow 0$, $\Delta \nu _{FWHM}/\nu _{line}\simeq
0.162$. The corresponding value shown in Table 1 is slightly
larger than this. It might be due to the width of the rest frame
line feature. When calculating the line described by (9) in the
case of $\Gamma =100$ by replacing $n_2=0$ with $n_2=-2$, we
really obtain $\Delta \nu _{FWHM}/\nu _{line}\simeq 0.162$.

\section{Effect on other lines}

We consider in the following the effect on other lines.

{\bf a) Weak narrow emission lines}

We take
\begin{equation}
\nu _{0,line,1}/\nu _{0,p}=0.01,\qquad h_1=10,\qquad n_1=-8
\end{equation}
and
\begin{equation}
\nu _{0,line,2}/\nu _{0,p}=100,\qquad h_2=0.01,\qquad n_2=0.
\end{equation}

\begin{figure}[tbp]
\vbox to 3.0in{\rule{0pt}{3.0in}} \includegraphics{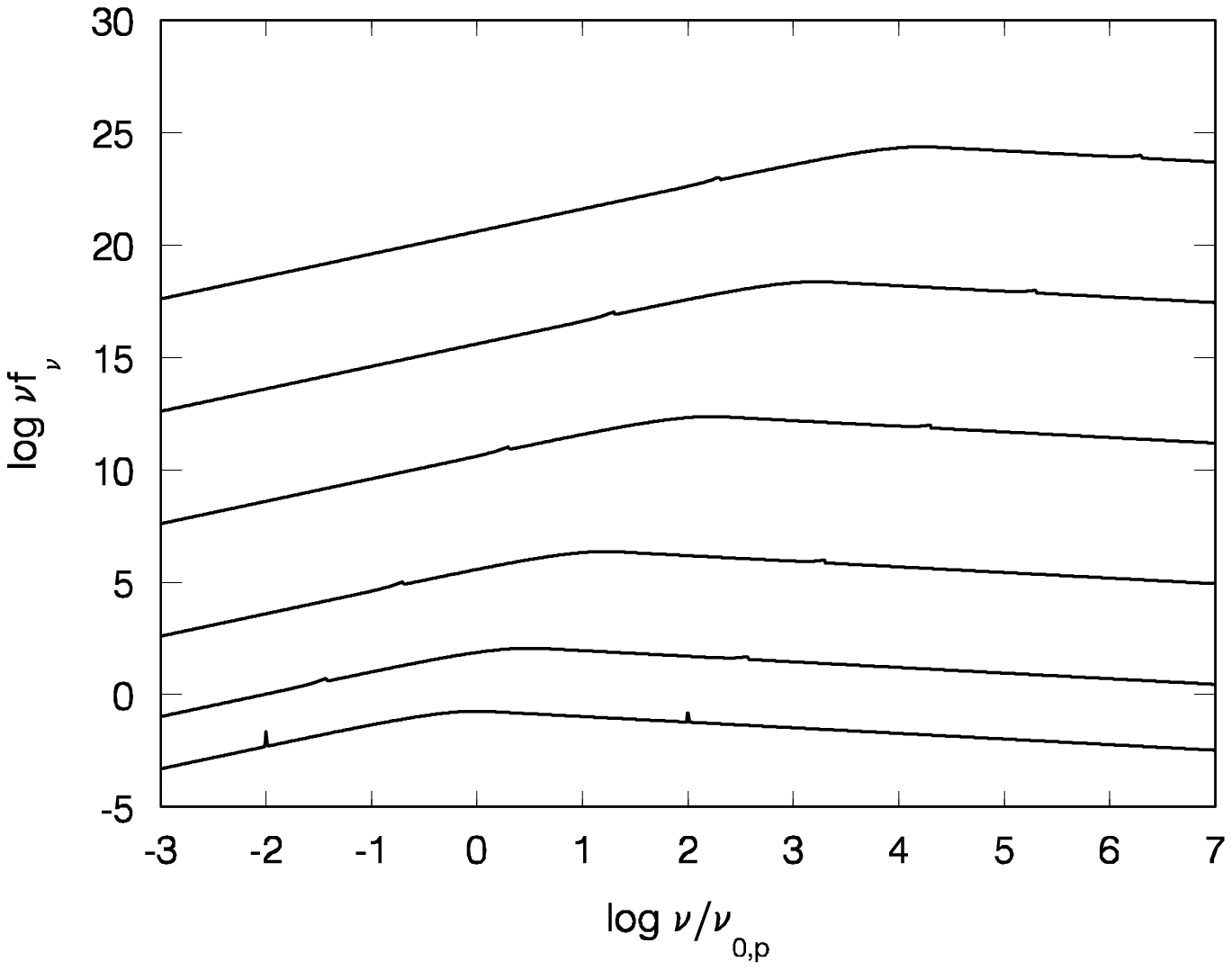} \caption{The
expected spectrum of a fireball with its rest frame radiation
containing the two weak emission lines defined by (10) and (11),
where other parameters and the symbols are the same as those
adopted in Fig. 1.} \label{Fig2}
\end{figure}

Shown in Fig. 2 are the $\log (\nu f_\nu )-\log (\nu /\nu _{0,p})$
curves with the two weak emission lines. The figure shows that, in
the spectrum of the fireball considered above, rest frame weak
narrow emission lines also shift to higher energy bands but are
significantly smoothed and would hardly be detectable.

{\bf b) Broad emission lines}

We take
\begin{equation}
\nu _{0,line,1}/\nu _{0,p}=0.01,\qquad h_1=10,\qquad n_1=-5
\end{equation}
and
\begin{equation}
\nu _{0,line,2}/\nu _{0,p}=100,\qquad h_2=0.01,\qquad n_2=3.
\end{equation}

\begin{figure}[tbp]
\vbox to 3.0in{\rule{0pt}{3.0in}} \includegraphics{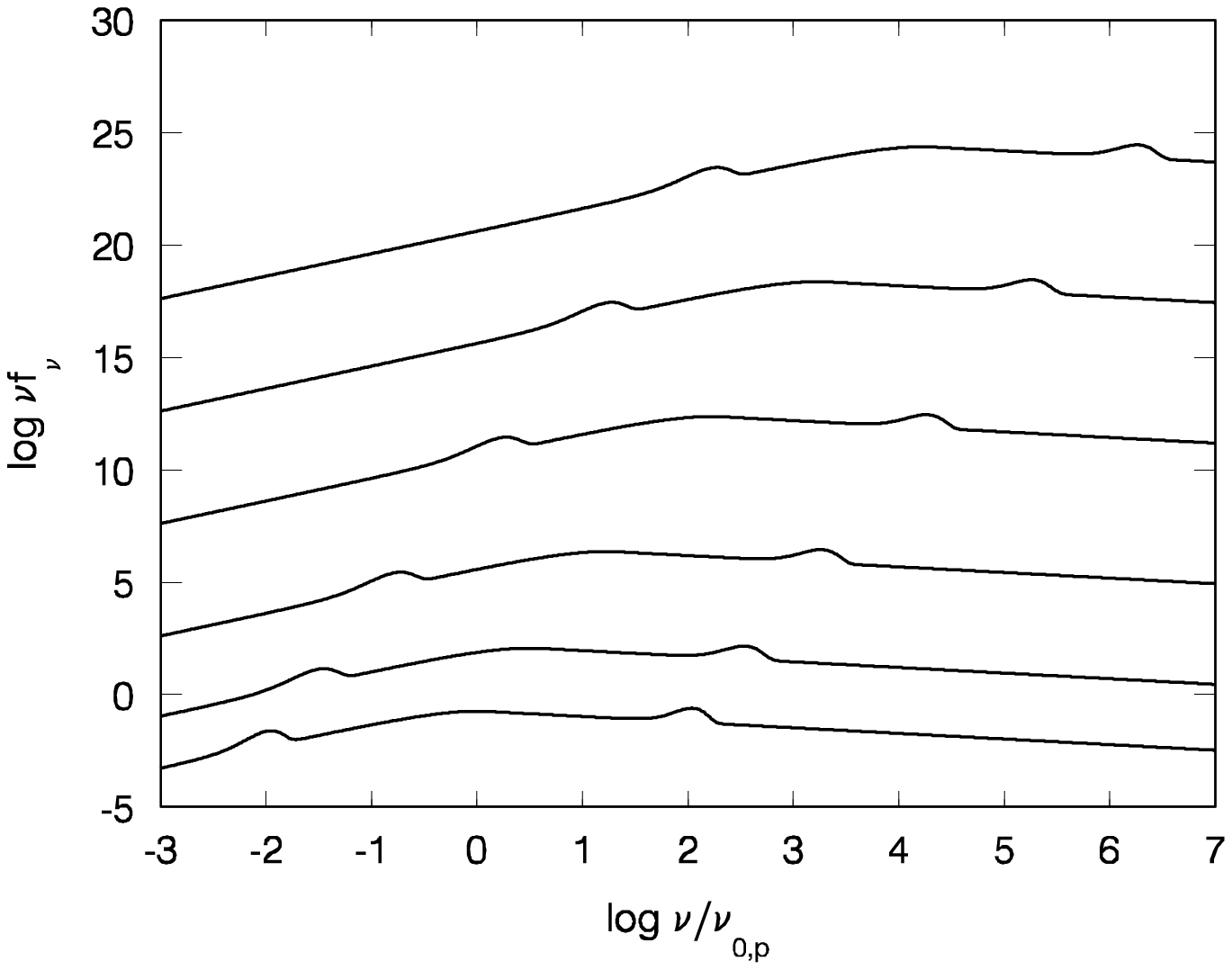} \caption{The
expected spectrum of a fireball with its rest frame radiation
containing the two broad emission lines defined by (12) and (13),
where other parameters and the symbols are the same as those
adopted in Fig. 1.} \label{Fig3}
\end{figure}

Shown in Fig. 3 are the $\log (\nu f_\nu )-\log (\nu /\nu _{0,p})$
curves with the two broad emission lines. It shows that rest frame
broad emission lines also shift to higher energy bands but their
features remain almost unchanged and therefore would be easier to
detect.

{\bf c) Narrow absorption lines}

We take
\begin{equation}
\nu _{0,line,1}/\nu _{0,p}=0.01,\qquad h_1=-1000,\qquad n_1=-8
\end{equation}
and
\begin{equation}
\nu _{0,line,2}/\nu _{0,p}=100,\qquad h_2=-1,\qquad n_2=0.
\end{equation}

\begin{figure}[tbp]
\vbox to 3.0in{\rule{0pt}{3.0in}} \includegraphics{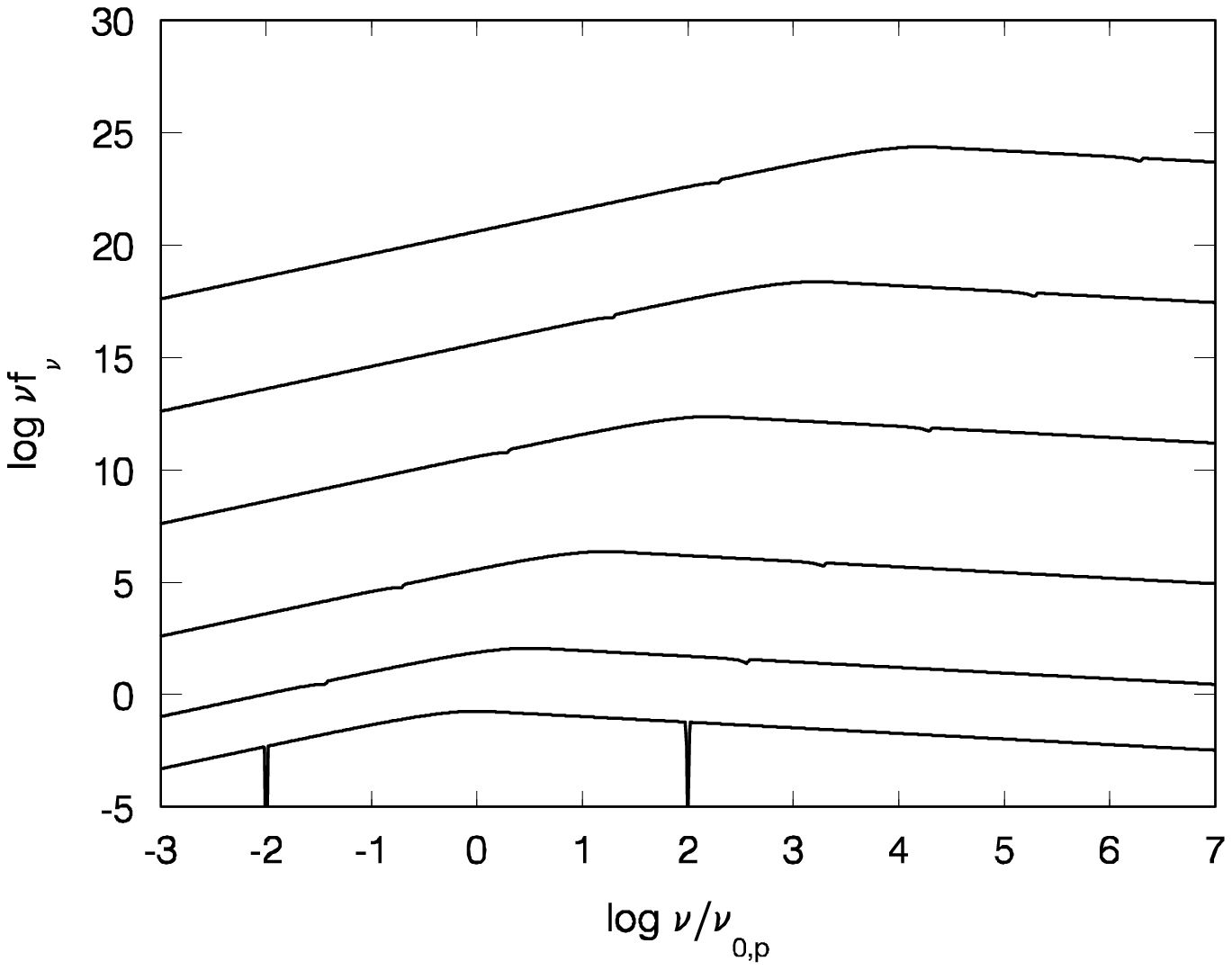} \caption{The
expected spectrum of a fireball with its rest frame radiation
containing the two narrow absorption lines defined by (14) and
(15), where other parameters and the symbols are the same as those
adopted in Fig. 1.} \label{Fig4}
\end{figure}

Shown in Fig. 4 are the $\log (\nu f_\nu )-\log (\nu /\nu _{0,p})$
curves with the two narrow absorption lines. It shows that rest
frame narrow absorption lines also shift to higher energy bands
but are significantly smoothed and would hardly be detectable.

{\bf d) Broad absorption lines}

We take
\begin{equation}
\nu _{0,line,1}/\nu _{0,p}=0.01,\qquad h_1=-10,\qquad n_1=-5
\end{equation}
and
\begin{equation}
\nu _{0,line,2}/\nu _{0,p}=100,\qquad h_2=-0.01,\qquad n_2=3.
\end{equation}

\begin{figure}[tbp]
\vbox to 3.0in{\rule{0pt}{3.0in}} \includegraphics{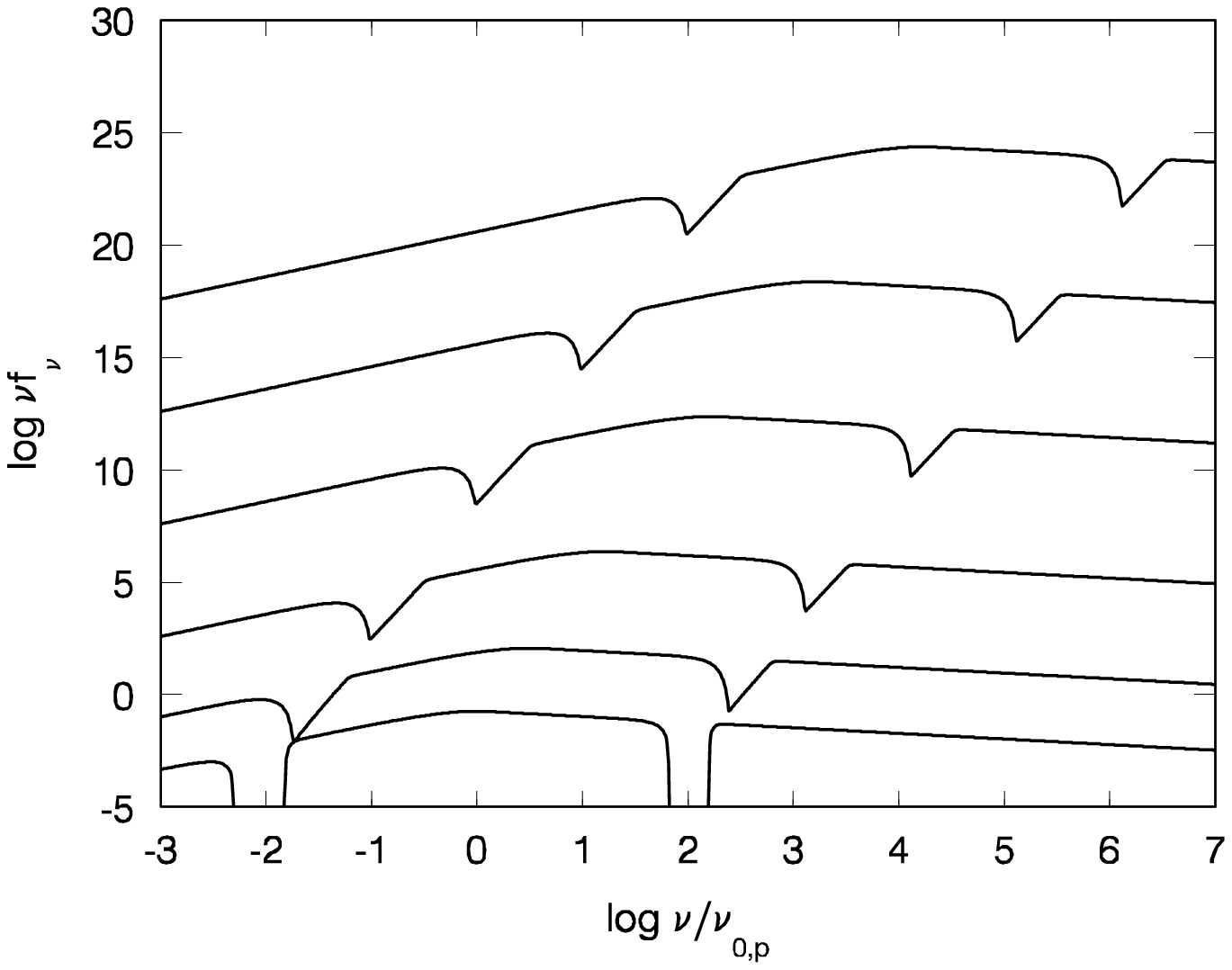} \caption{The
expected spectrum of a fireball with its rest frame radiation
containing the two broad absorption lines defined by (16) and
(17), where other parameters and the symbols are the same as those
adopted in Fig. 1.} \label{Fig5}
\end{figure}

Shown in Fig. 5 are the $\log (\nu f_\nu )-\log (\nu /\nu _{0,p})$
curves with the two broad absorption lines. It shows that rest
frame broad absorption lines also shift to higher energy bands but
are significantly smoothed so that deep gaps would be
significantly filled. The broad absorption features would remain
and therefore would be detectable.

{\bf e) Weak broad absorption lines}

We take
\begin{equation}
\nu _{0,line,1}/\nu _{0,p}=0.01,\qquad h_1=-2,\qquad n_1=-5
\end{equation}
and
\begin{equation}
\nu _{0,line,2}/\nu _{0,p}=100,\qquad h_2=-0.002,\qquad n_2=3.
\end{equation}

\begin{figure}[tbp]
\vbox to 3.0in{\rule{0pt}{3.0in}} \includegraphics{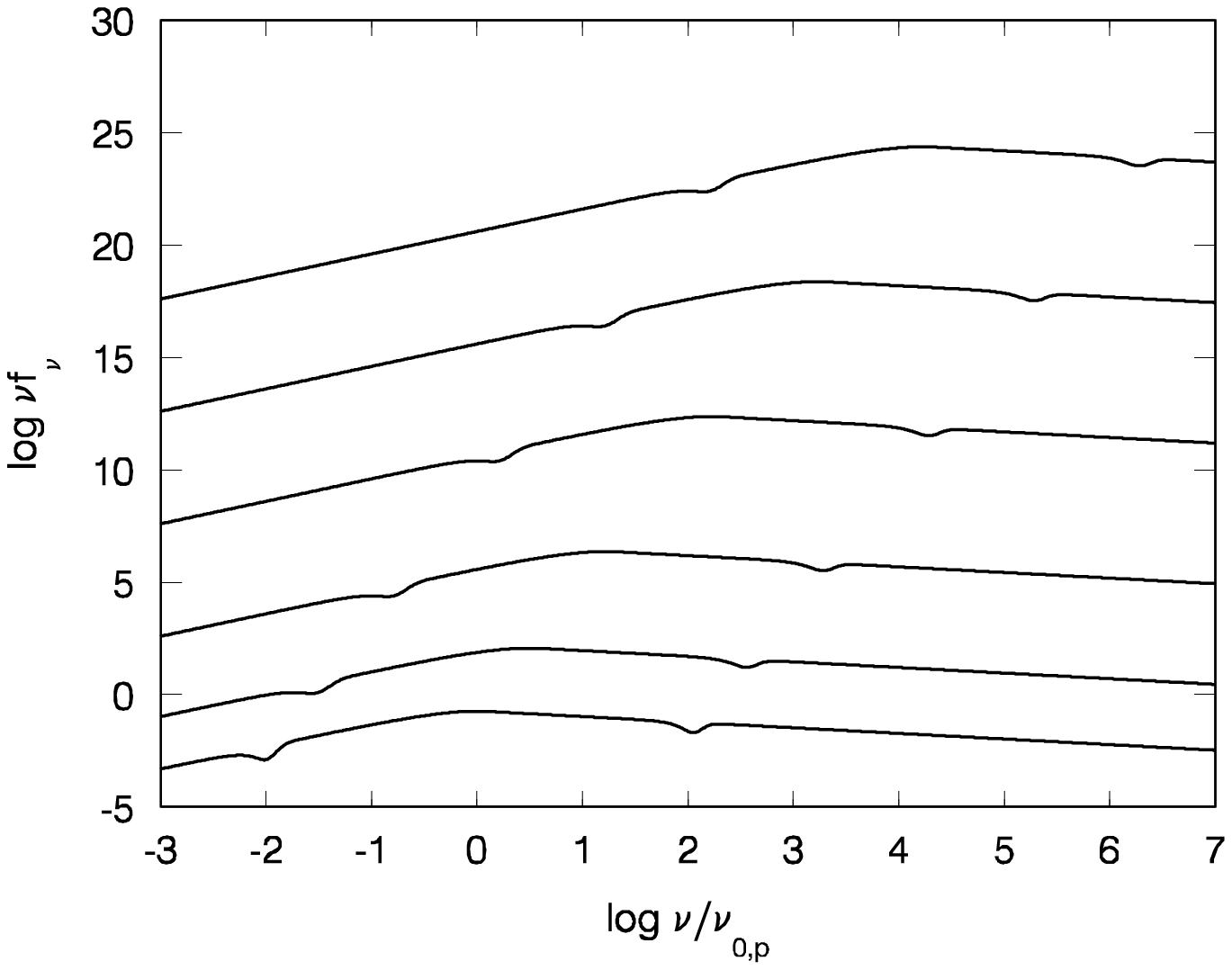} \caption{The
expected spectrum of a fireball with its rest frame radiation
containing the two weak broad absorption lines defined by (18) and
(19), where other parameters and the symbols are the same as those
adopted in Fig. 1.} \label{Fig6}
\end{figure}

Shown in Fig. 6 are the $\log (\nu f_\nu )-\log (\nu /\nu _{0,p})$
curves with the two weak broad absorption lines. It shows that
rest frame weak broad absorption lines also shift to higher energy
bands but their features remain almost unchanged and would be
detectable.

{\bf f) Emission line forest}

We consider two emission line forests at lower and higher energy
bands respectively. We take
\begin{equation}
\nu _{0,line,1}/\nu _{0,p}=0.01,\qquad h_1=1000,\qquad n_1=-8,
\end{equation}
\begin{equation}
\nu _{0,line,2}/\nu _{0,p}=0.012,\qquad h_2=1000,\qquad n_2=-8,
\end{equation}
\begin{equation}
\nu _{0,line,3}/\nu _{0,p}=0.014,\qquad h_3=1000,\qquad n_3=-8,
\end{equation}
\begin{equation}
\nu _{0,line,4}/\nu _{0,p}=0.016,\qquad h_4=1000,\qquad n_4=-8,
\end{equation}
and
\begin{equation}
\nu _{0,line,5}/\nu _{0,p}=100,\qquad h_5=1,\qquad n_5=0,
\end{equation}
\begin{equation}
\nu _{0,line,6}/\nu _{0,p}=120,\qquad h_6=1,\qquad n_6=0,
\end{equation}
\begin{equation}
\nu _{0,line,7}/\nu _{0,p}=140,\qquad h_7=1,\qquad n_7=0,
\end{equation}
\begin{equation}
\nu _{0,line,8}/\nu _{0,p}=160,\qquad h_8=1,\qquad n_8=0.
\end{equation}

\begin{figure}[tbp]
\vbox to 3.0in{\rule{0pt}{3.0in}} \includegraphics{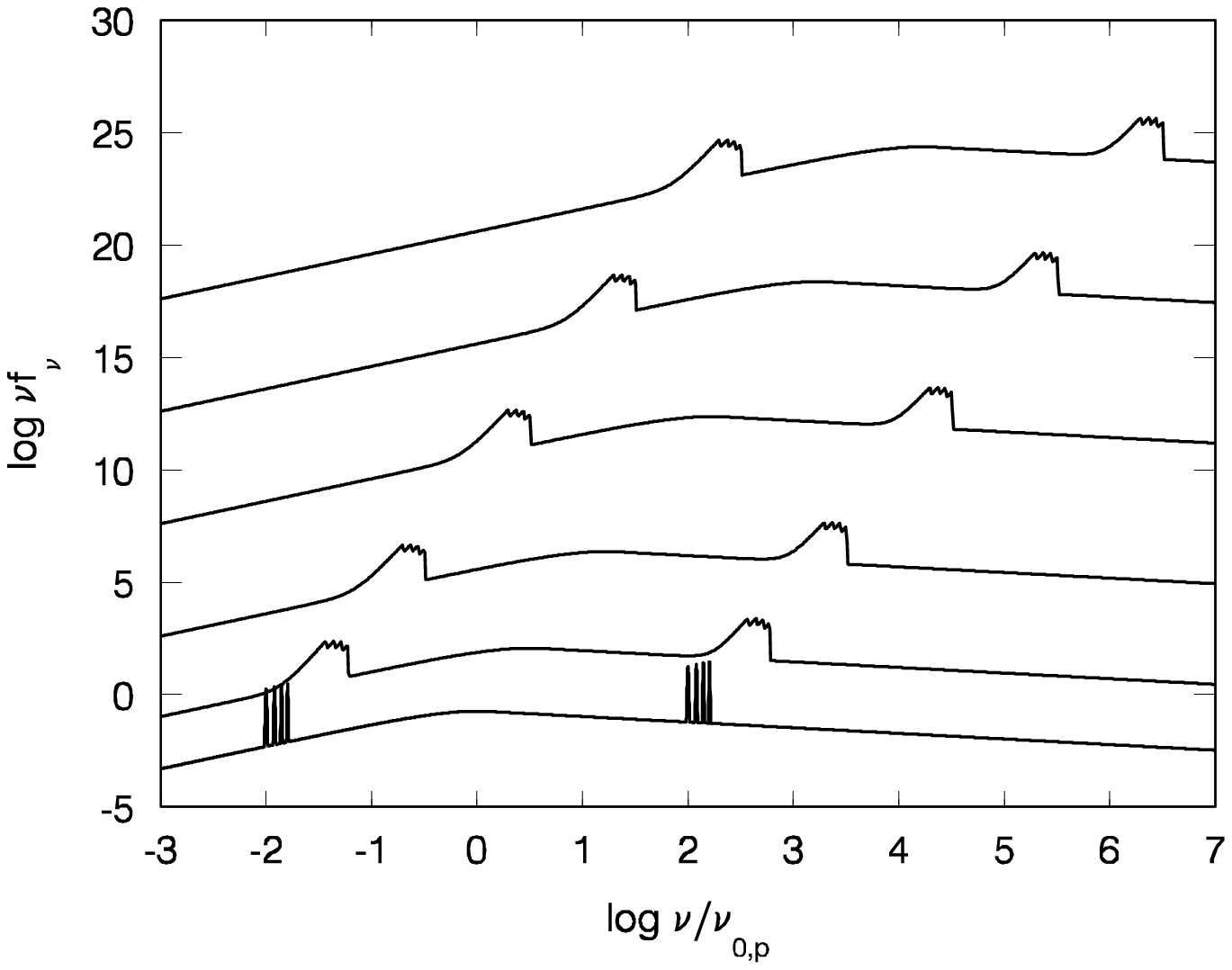} \caption{The
expected spectrum of a fireball with its rest frame radiation
containing the two emission line forests defined by (20)---(27),
where other parameters and the symbols are the same as those
adopted in Fig. 1.} \label{Fig7}
\end{figure}

Shown in Fig. 7 are the $\log (\nu f_\nu )-\log (\nu /\nu _{0,p})$
curves with the two emission line forests. It shows that rest
frame emission line forests also shift to higher energy bands but
they are significantly smoothed and each of them forms a single
broad line feature.

{\bf g) Absorption line forest}

We consider two absorption line forests at lower and higher energy
bands respectively and take
\begin{equation}
\nu _{0,line,1}/\nu _{0,p}=0.01,\qquad h_1=-1000,\qquad n_1=-8,
\end{equation}
\begin{equation}
\nu _{0,line,2}/\nu _{0,p}=0.012,\qquad h_2=-1000,\qquad n_2=-8,
\end{equation}
\begin{equation}
\nu _{0,line,3}/\nu _{0,p}=0.014,\qquad h_3=-1000,\qquad n_3=-8,
\end{equation}
\begin{equation}
\nu _{0,line,4}/\nu _{0,p}=0.016,\qquad h_4=-1000,\qquad n_4=-8,
\end{equation}
and
\begin{equation}
\nu _{0,line,5}/\nu _{0,p}=100,\qquad h_5=-1,\qquad n_5=0,
\end{equation}
\begin{equation}
\nu _{0,line,6}/\nu _{0,p}=120,\qquad h_6=-1,\qquad n_6=0,
\end{equation}
\begin{equation}
\nu _{0,line,7}/\nu _{0,p}=140,\qquad h_7=-1,\qquad n_7=0,
\end{equation}
\begin{equation}
\nu _{0,line,8}/\nu _{0,p}=160,\qquad h_8=-1,\qquad n_8=0.
\end{equation}

\begin{figure}[tbp]
\vbox to 3.0in{\rule{0pt}{3.0in}} \includegraphics{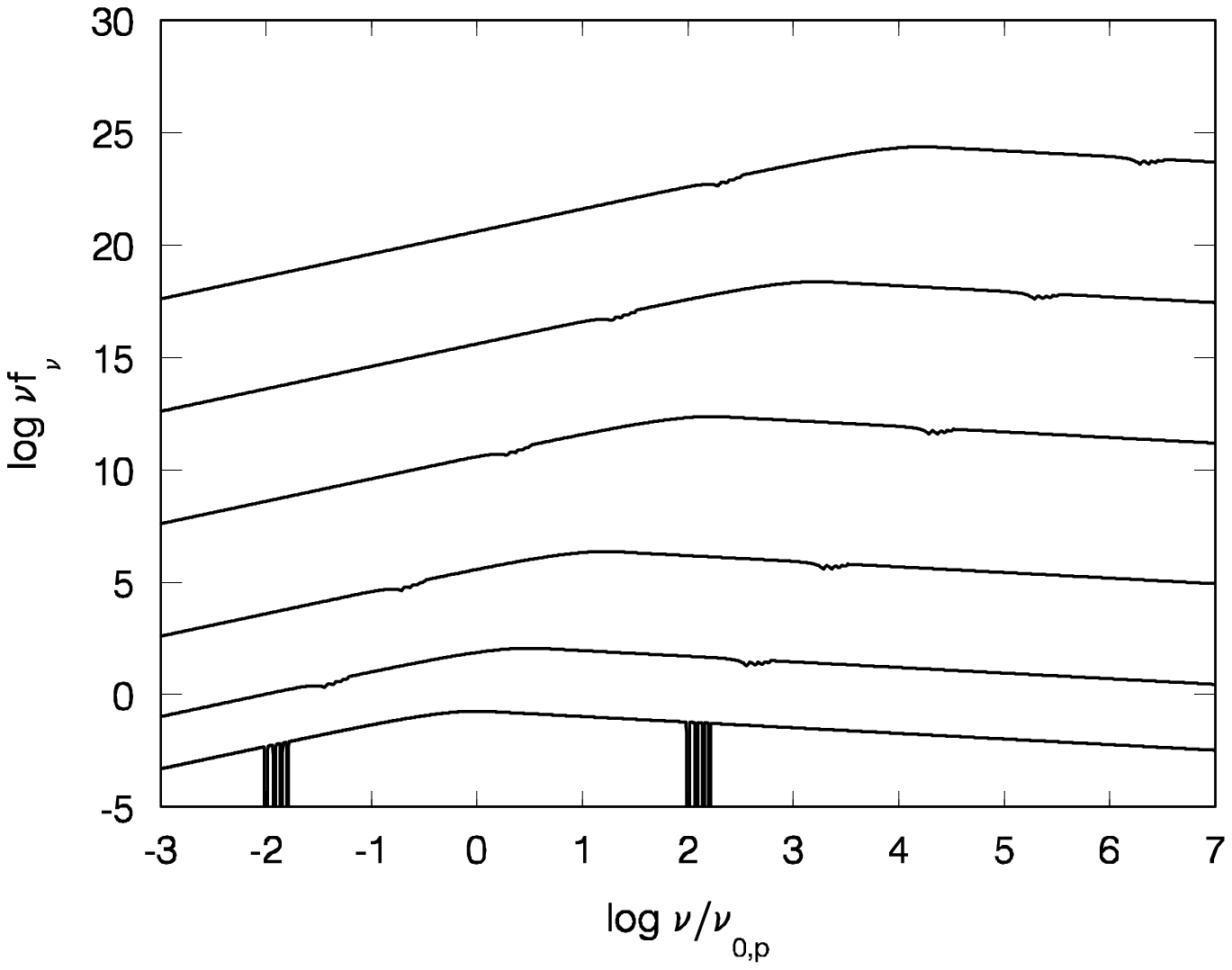} \caption{The
expected spectrum of a fireball with its rest frame radiation
containing the two absorption line forests defined by (28)---(35),
where other parameters and the symbols are the same as those
adopted in Fig. 1.} \label{Fig8}
\end{figure}

Shown in Fig. 8 are the $\log (\nu f_\nu )-\log (\nu /\nu _{0,p})$
curves with the two absorption line forests. It shows that rest
frame absorption line forests also shift to higher energy bands
but are significantly smoothed and would hardly be detectable.

\section{Effect of time dependence of the line intensity or variation of the
Lorentz factor}

Here we investigate the effect of time dependence of the line
intensity and the effect of variation of $\Gamma $.

{\bf a) The case of the line intensity varying with time}

We consider rest frame radiation containing a constant GRB form
and two time dependent intensity narrow lines. In this situation,
the formulas adopted in section 2 are valid when we replace $h_i$
in (7) with $ h_i(t_{0,\theta })$. Suppose the line intensity is
an exponential function of time:
\begin{equation}
h_i(t_{0,\theta })=h_{i,0}\exp (-\frac{t_{0,\theta }-t_{0,c}}{\tau _0}%
)\qquad \qquad \qquad \qquad (t_{0,\theta }\geq t_{0,c}),
\end{equation}
where $h_{i,0}$, $t_{0,c}$ and $\tau _0$ are constants. As shown
in Paper I, $t_{0,\theta }$ and $t$ can be related by
\begin{equation}
t_{0,\theta }=\frac{t-t_c-D/c+(R_c/c)\cos \theta }{\Gamma (1-\beta
\cos \theta )}+t_{0,c},
\end{equation}
where $t_c$ and $R_c$ are constants. We take
\begin{equation}
\tau _0=100R_c/c.
\end{equation}
Therefore,
\begin{equation}
h_i(t_{0,\theta })=h_{i,0}\exp [-\frac{\frac{c(t-t_c)-D}{R_c}+\cos \theta }{%
100\Gamma (1-\beta \cos \theta )}]\qquad
[\frac{t-t_c-D/c+(R_c/c)\cos \theta }{\Gamma (1-\beta \cos \theta
)}\geq 0].
\end{equation}
We consider different observation times and take
\begin{equation}
t_1=\frac Dc+t_c,
\end{equation}
\begin{equation}
t_2=\frac{R_c+D}c+t_c
\end{equation}
and
\begin{equation}
t_3=\frac{10R_c+D}c+t_c.
\end{equation}
Once more we adopt (8) and (9), where $h_i$ stands for $h_{i,0}$
($i=1,2$).

\begin{figure}[tbp]
\vbox to 3.0in{\rule{0pt}{3.0in}} \includegraphics{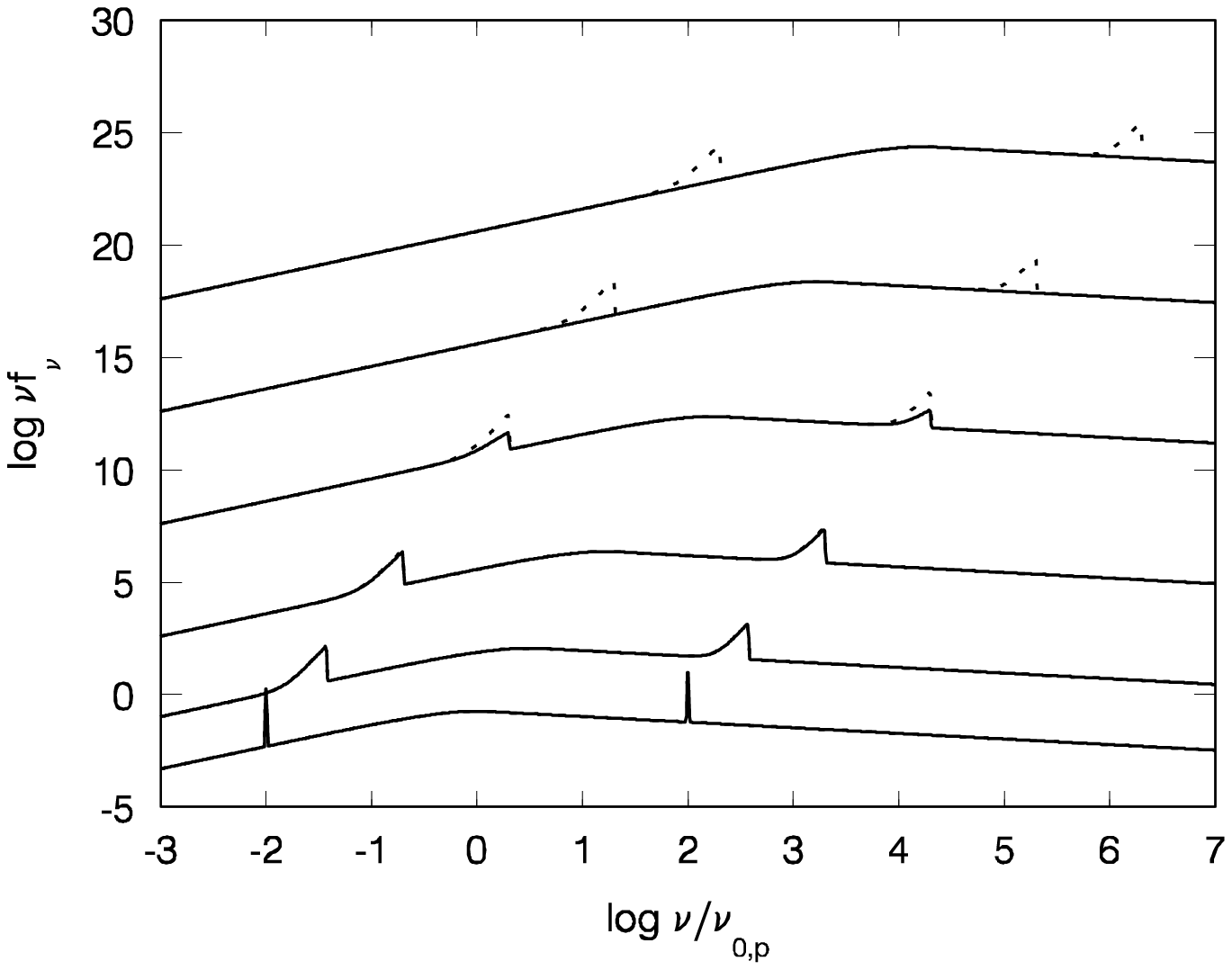} \caption{The
expected spectrum of a fireball containing in its rest frame
radiation the two time dependent intensity emission lines
[confined by (39)] for observation time $t_1$ [defined by (40)],
where we take $2\pi I_0\nu _{0,p}\widetilde{R}^2(t_1)/D^2=1$. The
solid lines from the bottom to the top correspond to $\Gamma
=1,2,10,100,1000,10000$, respectively, and the dotted lines (some
are totally overlapped by the corresponding solid lines) are those
presented in Fig. 1.} \label{Fig9}
\end{figure}

\begin{figure}[tbp]
\vbox to 3.0in{\rule{0pt}{3.0in}} \includegraphics{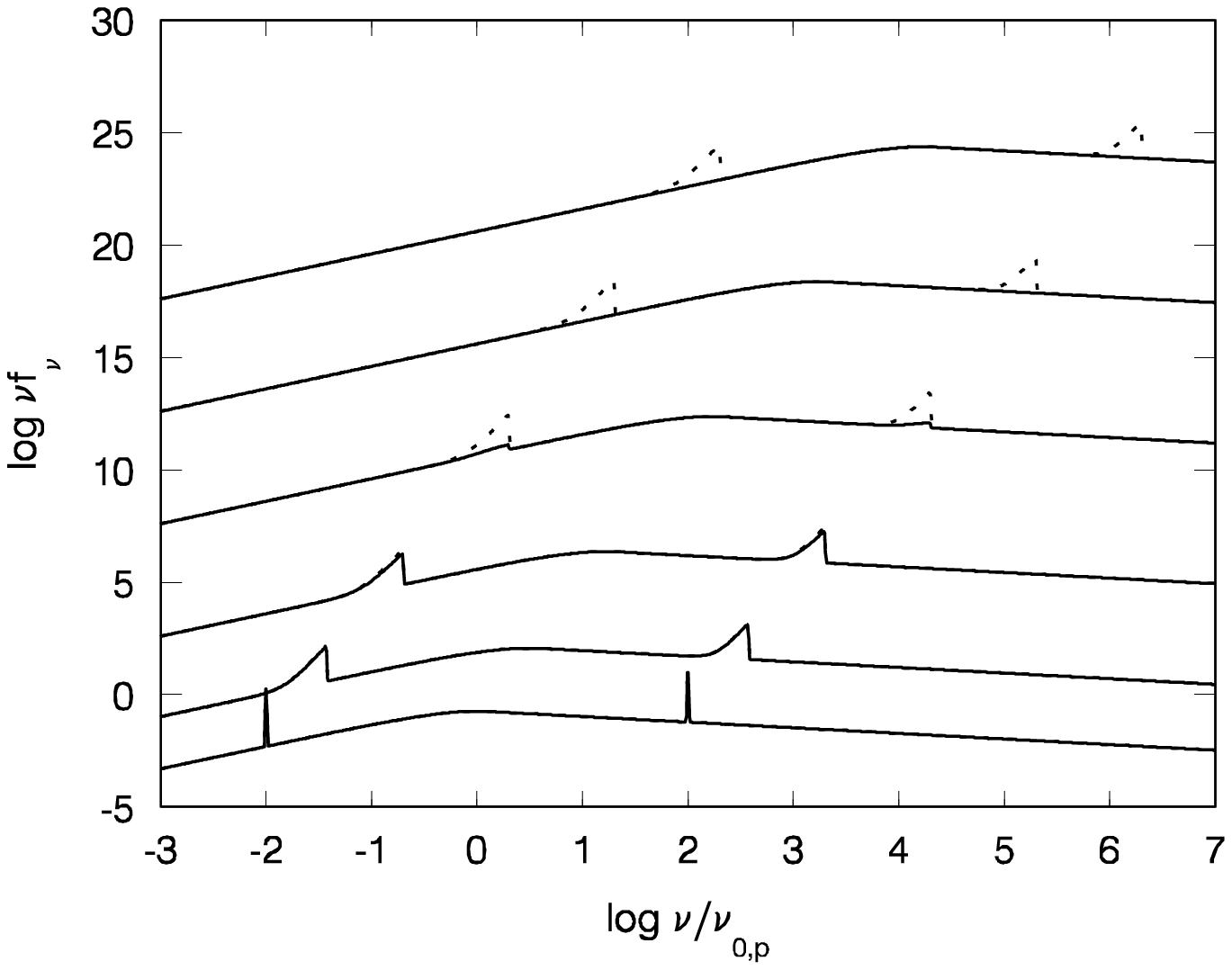} \caption{The
expected spectrum of a fireball containing in its rest frame
radiation the two time dependent intensity emission lines for
observation
time $t_2$ [defined by (41)], where we take $2\pi I_0\nu _{0,p}\widetilde{R}%
^2(t_2)/D^2=1$. Other parameters and the symbols are the same as
those adopted in Fig. 9.} \label{Fig10}
\end{figure}

\begin{figure}[tbp]
\vbox to 3.0in{\rule{0pt}{3.0in}} \includegraphics{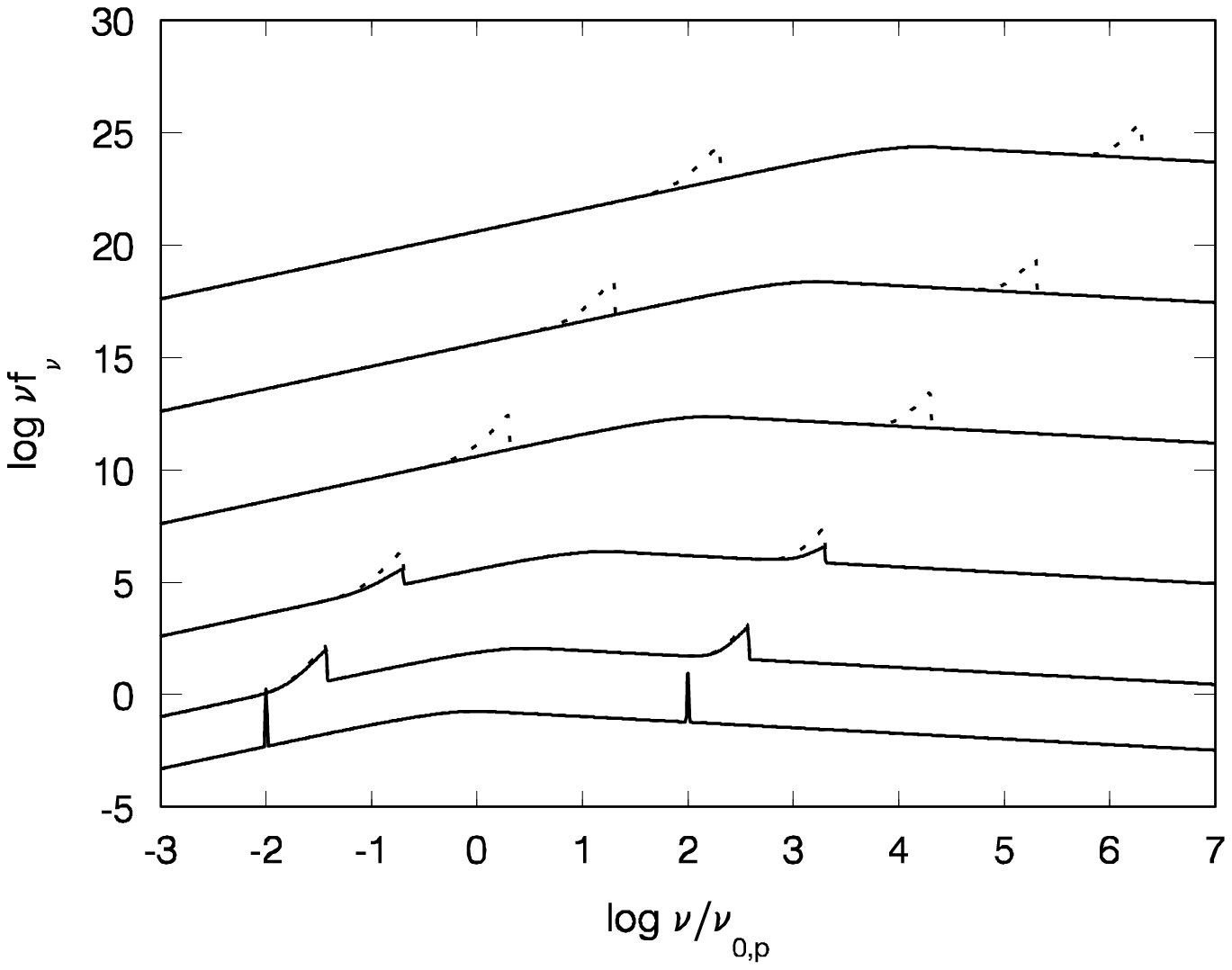} \caption{The
expected spectrum of a fireball containing in its rest frame
radiation the two time dependent intensity emission lines for
observation time $t_3$ [defined by (42)], where we take $2\pi
I_0\nu _{0,p}\widetilde{R} ^2(t_3)/D^2=1$. Other parameters and
the symbols are the same as those adopted in Fig. 9.}
\label{Fig11}
\end{figure}

The plots of $\log (\nu f_\nu )-\log (\nu /\nu _{0,p})$ for the
fireball containing in its rest frame radiation the two time
dependent intensity emission lines for the three observation times
are shown by Figs. 9--11 respectively, where various values of
$\Gamma $ are considered for each
figure. For the sake of comparison, we take $2\pi I_0\nu _{0,p}\widetilde{R}%
^2(t_1)/D^2=1$, $2\pi I_0\nu _{0,p}\widetilde{R}^2(t_2)/D^2=1$ and
$2\pi I_0\nu _{0,p}\widetilde{R}^2(t_3)/D^2=1$ for the three
figures, and the curves shown in Fig. 1. are also plotted.

Similar to that shown in Fig. 1, in the spectrum of a
relativistically expanding fireball, rest frame dimming narrow
emission lines would also shift to higher energy bands and would
be significantly broadened.
However, for fireballs with a very large Lorentz factor (say, $\Gamma >1000$%
), the line feature would disappear. As time progresses, the
feature would eventually disappear and this would happen at a much
earlier time for those fireballs with a relatively larger value of
the Lorentz factor.

{\bf b) The case of variation of the Lorentz factor}

Here we study the effect of variation of the Lorentz factor on the
spectrum of fireballs. The expected spectrum of fireballs
influenced by this variation is studied in Appendix B. The effect
would be shown more clearly when the influenced spectrum is
compared with the previously discussed one. Therefore, we consider
the same rest frame constant radiation discussed in section 2,
with the variation of the Lorentz factor being the only
difference. To employ what we derived in Appendix B, we assume
that the radiation is independent of direction, $I_{0,\nu
}(t_{0,\theta },\nu _{0,\theta },\theta )=I_{0,\nu }(t_{0,\theta
},\nu _{0,\theta })$, and there is no limit for the radiated
frequency and the radiation concerned lasts a sufficiently large
interval of time so that the integral limits of (B.15) are $\theta
_{\min }=0$ and $\theta _{\max }=\pi /2$ (see Paper I for a
detailed discussion). Applying (3), (B.7) and (B.16), we obtain
from (B.15) that
\begin{equation}
\nu f_\nu (t)=\frac{2\pi I_0\nu }{D^2}\int_0^{\pi
/2}\frac{R^2(t_\theta )g_{0,\nu }(\nu _{0,\theta })}{\Gamma
^3(t_\theta )[1-\beta (t_\theta )\cos \theta ]^3}\cos \theta \sin
\theta d\theta ,
\end{equation}
where we take $I_0(t_{0,\theta })=I_0$, and $g_{0,\nu }(\nu
_{0,\theta })$ is determined by (6). Frequencies $\nu _{0,\theta
}$ and $\nu $ are related by the Doppler effect [see (B.7)].

It is known that, in its deceleration phase, the Lorentz factor of
a fireball would develop with time following $t^{-3/8}$ in the
case of the adiabatic hydrodynamics while following $t^{-3/7}$ in
the case of the radiative hydrodynamics (Piran 1999). To
illustrate the effect, we consider the latter case and assume
\begin{equation}
\Gamma (t_\theta )=\max \{kt_\theta ^{-3/7},1\}\qquad \qquad
\qquad \qquad (t_\theta \geq t_c),
\end{equation}
where $k$ is a constant. Here we assign $t_c$ to be the time when
the deceleration starts and assign $\Gamma =1$ when the
deceleration as well as the expansion stop. Suppose $\Gamma _c>1$
is the value of the Lorentz factor of the fireball at time $t_c$,
then one obtains
\begin{equation}
k=\Gamma _ct_c^{3/7}.
\end{equation}

When $\Gamma (t_\theta )$ is known, the relation between the
coordinate time, $t_\theta $, of the differential surface of the
fireball concerned and the observation time, $t$, will be well
established, and the radius of the fireball as a function of time
will be determined (see Appendix B). From (B.9) we get
\begin{equation}
t_\theta -[\int_{t_c}^{t_\theta }\sqrt{1-\frac 1{(\max \{kt_\theta
^{-3/7},1\})^2}}dt_\theta ]\cos \theta =t+\frac{R_c}c\cos \theta
-\frac Dc.
\end{equation}

We find that, once $t$ is assigned, $t_\theta $ will be determined
by (46).
Then $\Gamma (t_\theta )$ will be determined by (44). Once both $t$ and $%
t_\theta $ are known, $R(t_\theta )$ will be easily determined. We
get from (B.8) and (46) that
\begin{equation}
R(t_\theta )=\frac{D-c(t-t_\theta )}{\cos \theta
}=R_c+c\int_{t_c}^{t_\theta }\sqrt{1-\frac 1{(\max \{kt_\theta
^{-3/7},1\})^2}}dt_\theta .
\end{equation}

To show the effect, different observation times should be
concerned. Let
\begin{equation}
p\equiv \frac{c(t-t_c)-D}{R_c}.
\end{equation}
Then
\begin{equation}
t=t_c+\frac{D+pR_c}c.
\end{equation}
We then come to
\begin{equation}
t_\theta -[\int_{t_c}^{t_\theta }\sqrt{1-\frac 1{(\max \{kt_\theta
^{-3/7},1\})^2}}dt_\theta ]\cos \theta =t_c+\frac{R_c}c(p+\cos
\theta )
\end{equation}
and
\begin{equation}
R(t_\theta )=\frac{R_c}{\cos \theta }(\frac{t_\theta
-t_c}{R_c/c}-p).
\end{equation}

To focus on the effect of variation of the Lorentz factor, we
require that, all photons observed at the assigned time must be
those emitted after the deceleration starts and before the
expansion stops. The first requirement leads to $p\geq 0$ (see
Appendix C). To meet the second requirement, $k$ cannot be
arbitrarily taken as long as $p$, $\Gamma _c$ and $R_c$ are fixed.
For given values of $p$, $\Gamma _c$ and $R_c$, there will a lower limit of $%
k$ (see Appendix C), i.e.,
\begin{equation}
k>k_0,
\end{equation}
with
\begin{equation}
k_0\equiv \Gamma _c[\frac{R_c(1+p)}{c(1-\beta _c)(\Gamma
_c^{7/3}-1)}]^{3/7},
\end{equation}
where $\beta _c\equiv \sqrt{\Gamma _c^2-1}/\Gamma _c$.

To show the development of the spectrum, we consider two different
observation times and take $p=(0,10)$. For the sake of comparison, we adopt $%
R_c/c=1$ and take $2\pi I_0\nu _{0,p}R_c^2/D^2=1$ and
$k=10k_0(p=10,\Gamma _c,R_c)$ for all situations concerned.

\begin{figure}[tbp]
\vbox to 3.0in{\rule{0pt}{3.0in}} \includegraphics{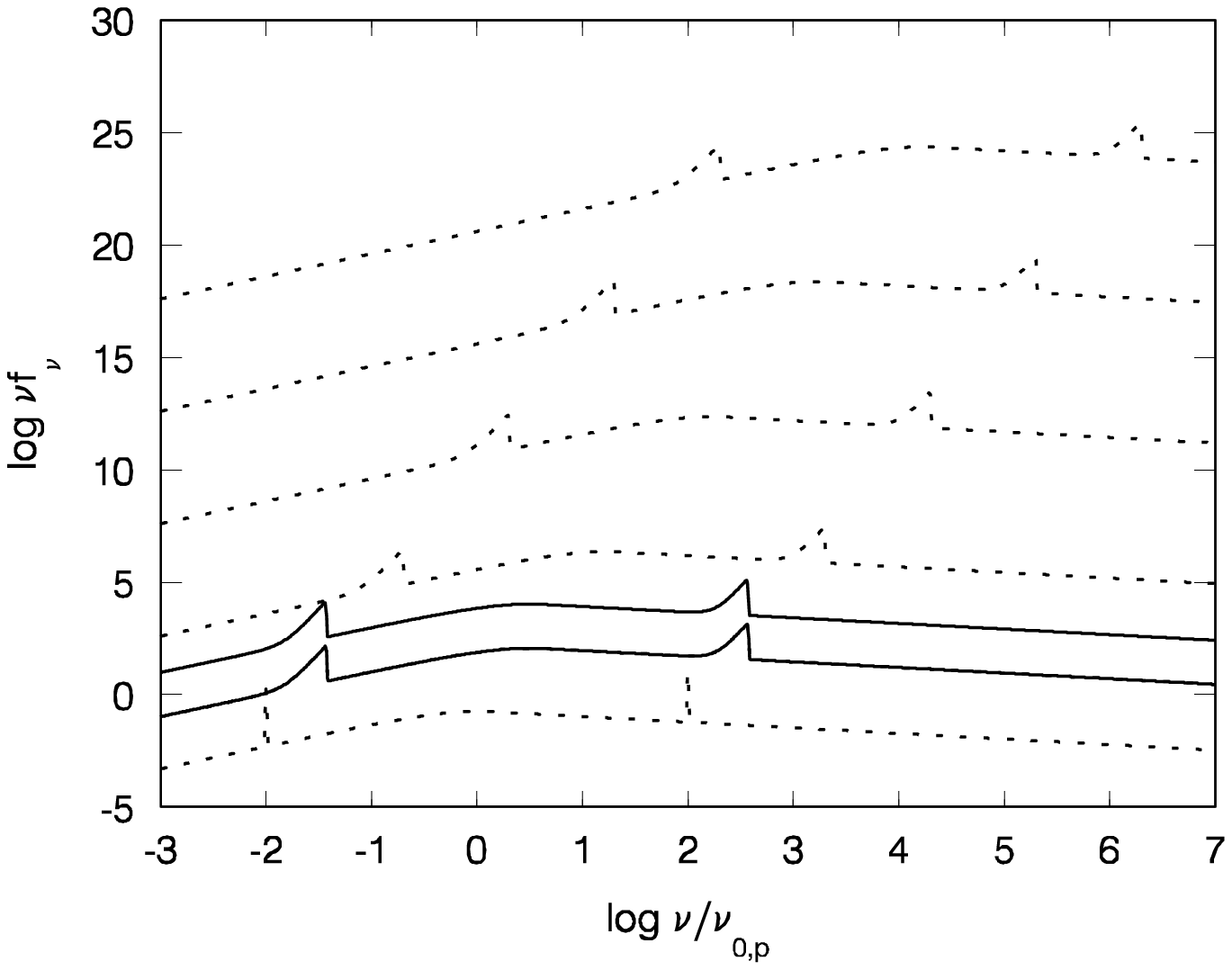} \caption{The
expected spectrum of a fireball with a variable Lorentz factor,
where the initial Lorentz factor is $\Gamma _c=2$ and the
radiation concerned is that discussed in section 2. We take
$R_c/c=1$, $2\pi I_0\nu _{0,p}R_c^2/D^2=1$, and
$k=10k_0(p=10,\Gamma _c=2,R_c/c=1)$. The solid lines from the
bottom to the top correspond to $p=0,10$, respectively, and the
dotted lines (some are totally overlapped by the corresponding
solid lines) are those presented in Fig. 1.} \label{Fig12}
\end{figure}

\begin{figure}[tbp]
\vbox to 3.0in{\rule{0pt}{3.0in}} \includegraphics{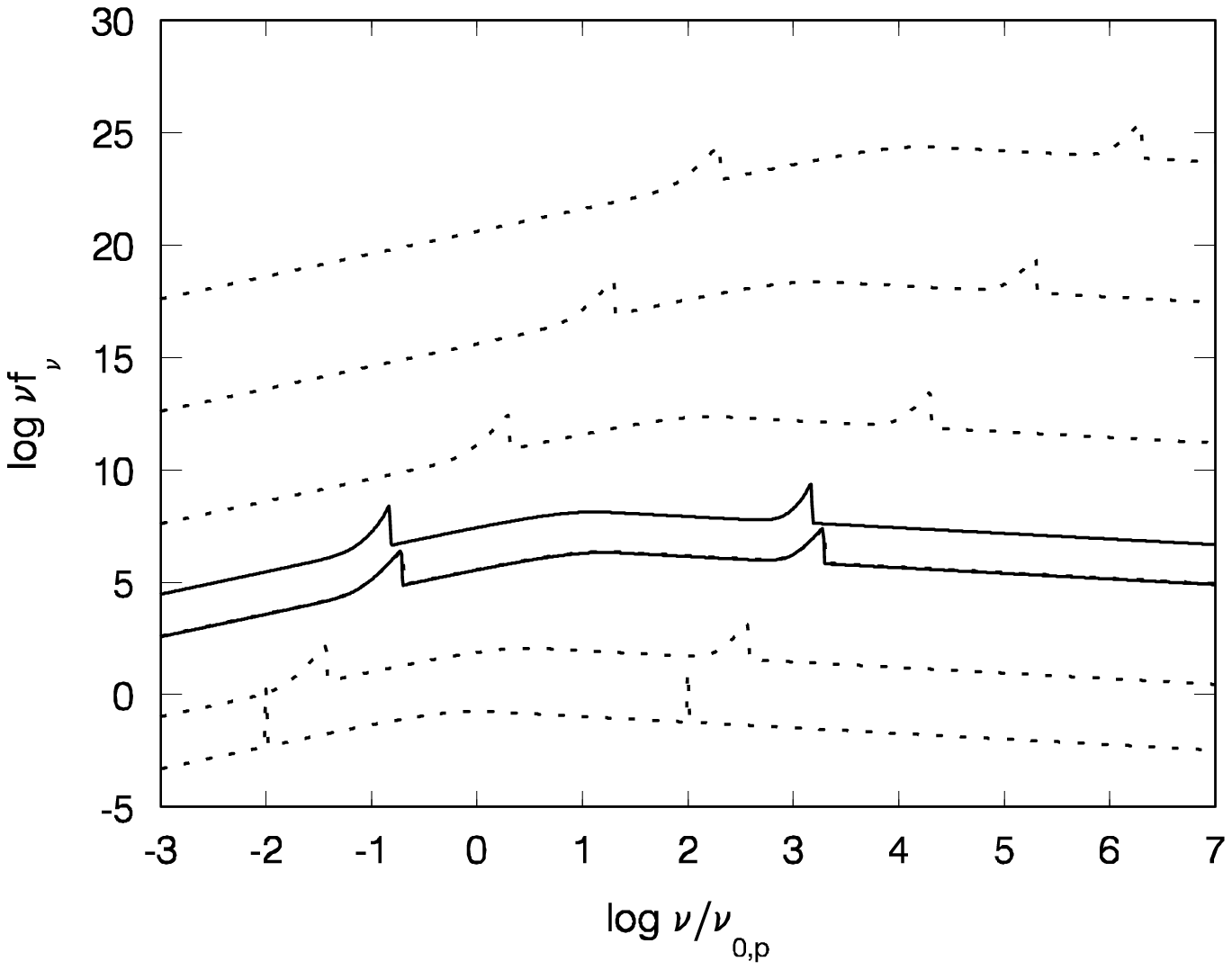} \caption{The
expected spectrum of a fireball with a variable Lorentz factor,
where $\Gamma _c=10$ and $k=10k_0(p=10,\Gamma _c=10,R_c/c=1)$.
Other parameters and the symbols are the same as those adopted in
Fig. 12.} \label{Fig13}
\end{figure}

\begin{figure}[tbp]
\vbox to 3.0in{\rule{0pt}{3.0in}} \includegraphics{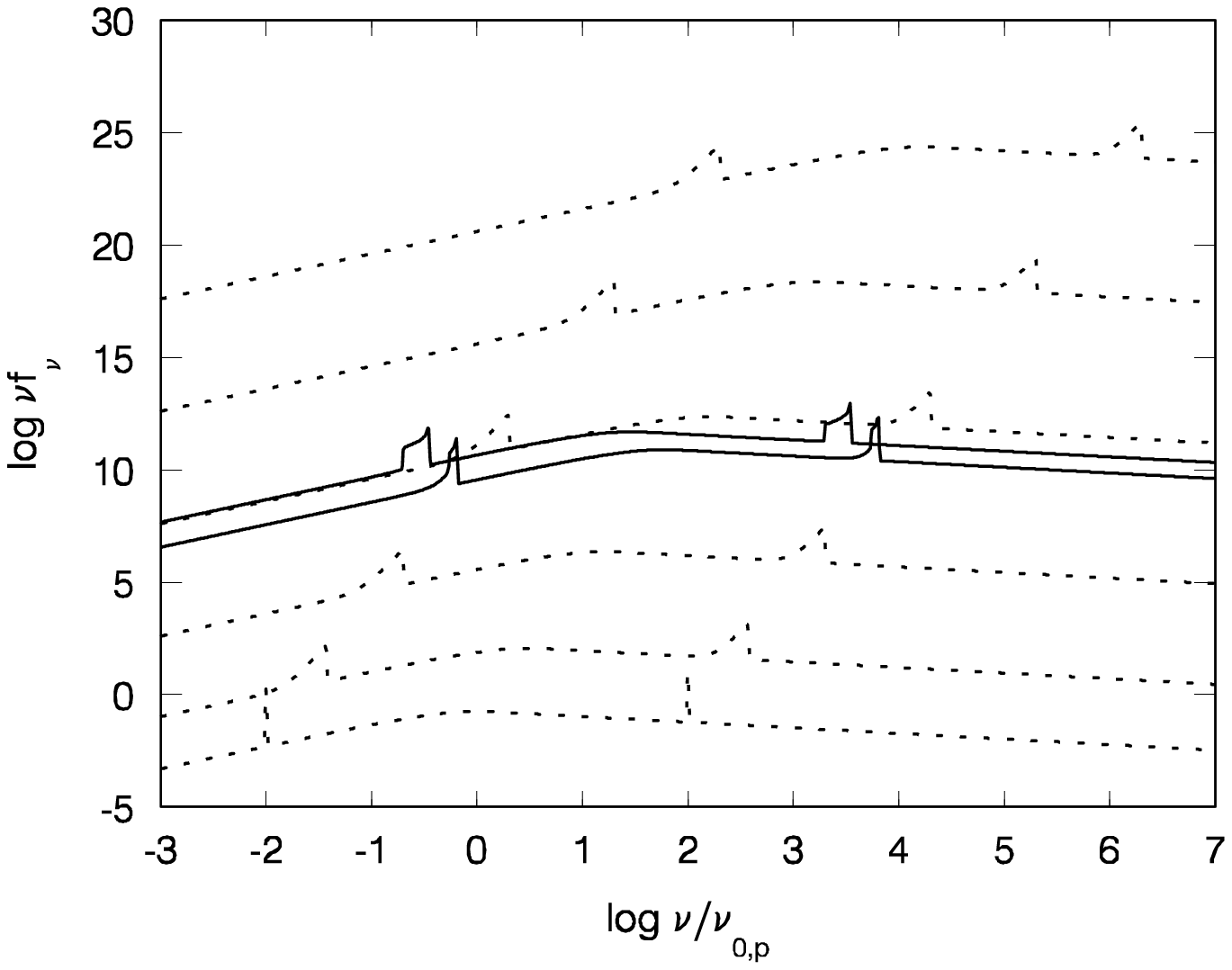} \caption{The
expected spectrum of a fireball with a variable Lorentz factor,
where $\Gamma _c=100$ and $k=10k_0(p=10,\Gamma _c=100,R_c/c=1)$.
Other parameters and the symbols are the same as those adopted in
Fig. 12.} \label{Fig14}
\end{figure}

\begin{figure}[tbp]
\vbox to 3.0in{\rule{0pt}{3.0in}} \includegraphics{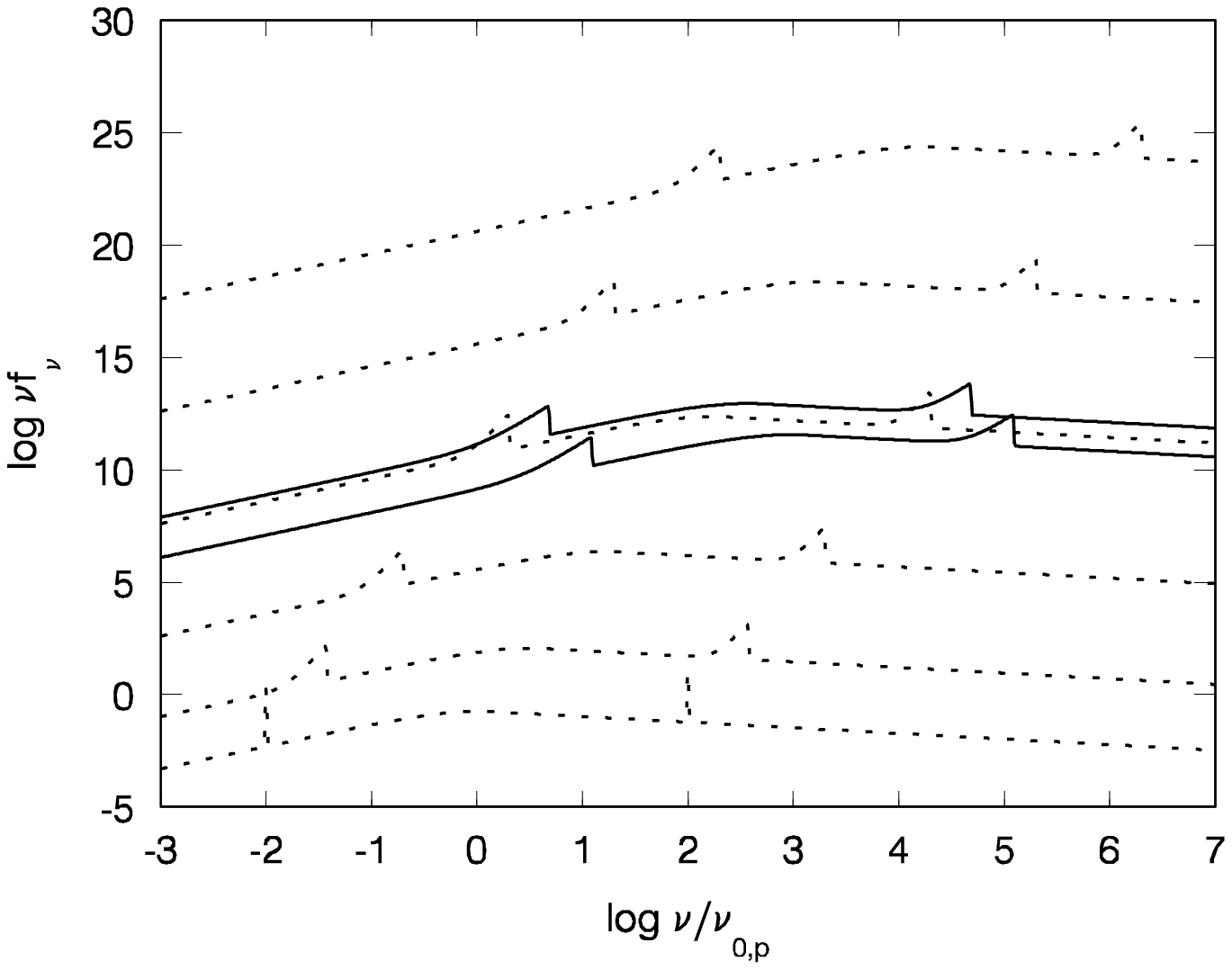} \caption{The
expected spectrum of a fireball with a variable Lorentz factor,
where $\Gamma _c=1000$ and $k=10k_0(p=10,\Gamma _c=1000,R_c/c=1)$.
Other parameters and the symbols are the same as those adopted in
Fig. 12.} \label{Fig15}
\end{figure}

\begin{figure}[tbp]
\vbox to 3.0in{\rule{0pt}{3.0in}} \includegraphics{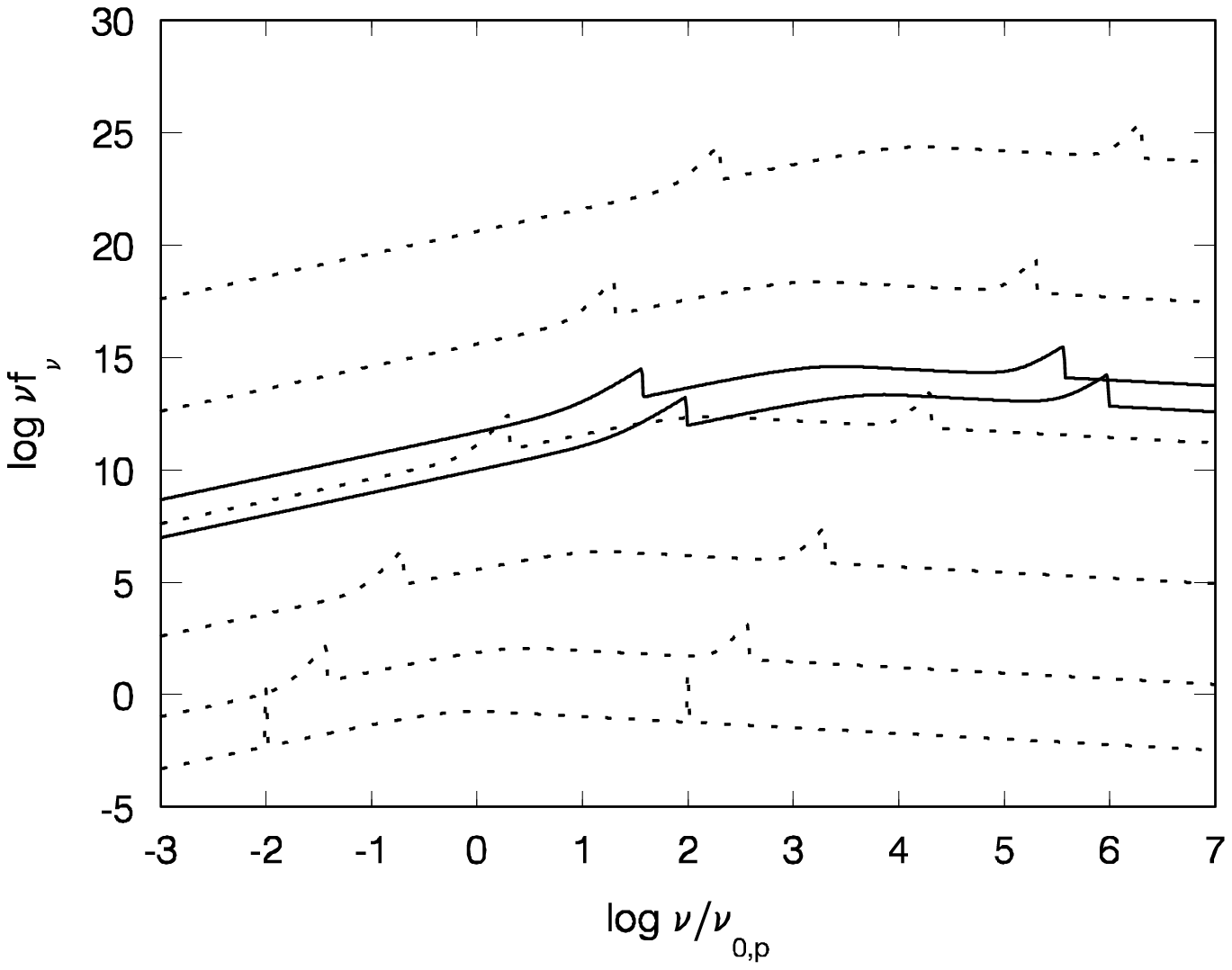} \caption{The
expected spectrum of a fireball with a variable Lorentz factor,
where $\Gamma _c=10000$ and $k=10k_0(p=10,\Gamma
_c=10000,R_c/c=1)$. Other parameters and the symbols are the same
as those adopted in Fig. 12.} \label{Fig16}
\end{figure}

The results are displayed in Figs. 12--16. All these figures show
that the flux observed at a relatively later time would be larger
than that observed at an earlier time, suggesting that, in the
situation studied here, the enhancement of the radius of the
fireball plays an important role in the magnitude of the flux.

Shown in Fig. 12 we find that, when $\Gamma _c$ is very small, as
the variation of the Lorentz factor is then not obvious, the
effect of deceleration would be insignificant. The positions of
emission lines would remain almost unchanged.

Figs. 14--16 reveal that the flux affected by the deceleration
would become much smaller than the constant radiation flux when
the initial Lorentz factor is large enough (say, $\Gamma _c\geq
100$), and the change of the positions of emission lines due to
the decrease of the Lorentz factor would be quite significant (the
positions of the lines would shift to lower energy bands as time
progresses).

\begin{figure}[tbp]
\vbox to 3.0in{\rule{0pt}{3.0in}} \includegraphics{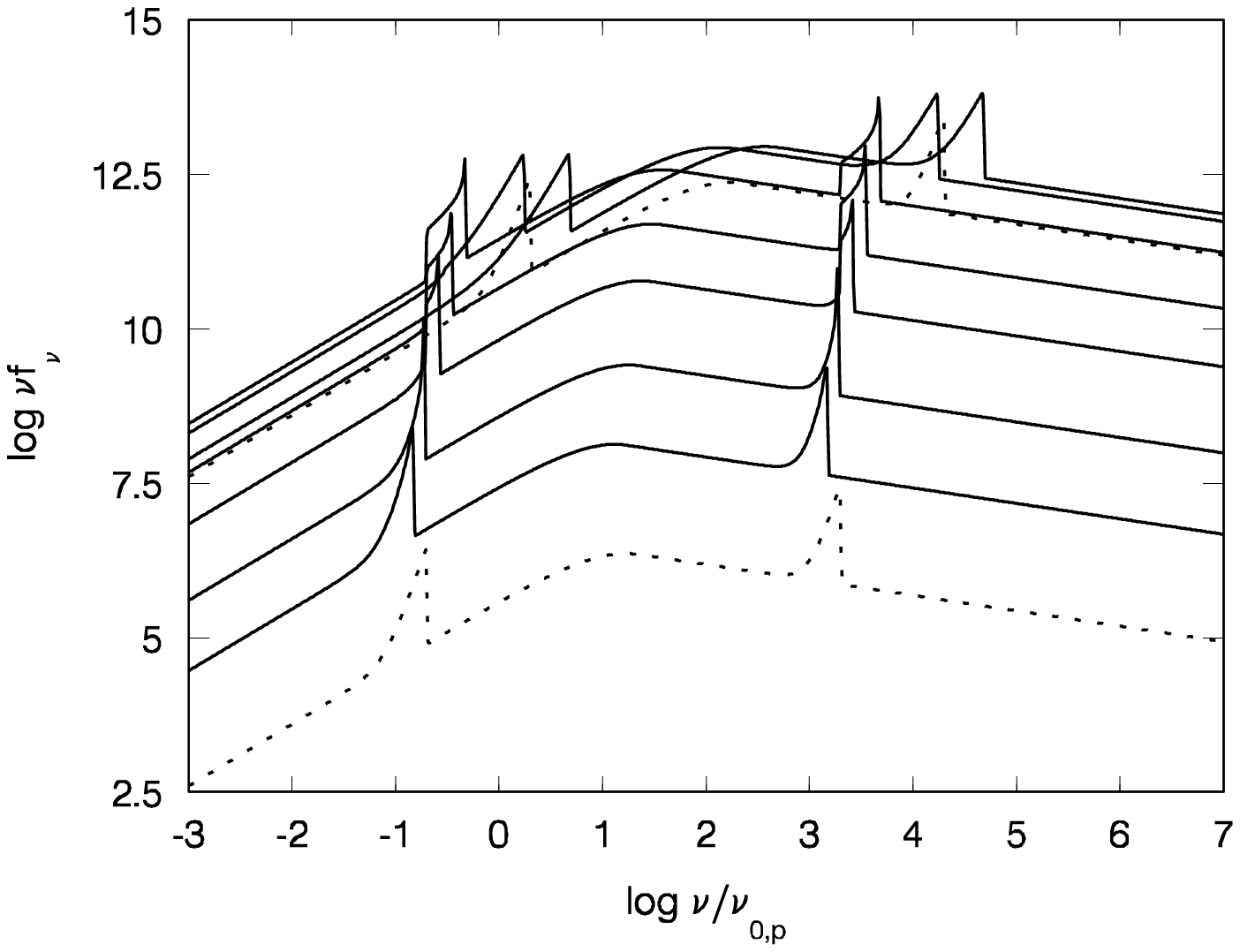} \caption{The
expected spectrum of a fireball with a variable Lorentz factor,
where we take $p=10$ and $k=10k_0(p=10,\Gamma _c,R_c/c=1)$ and
consider different values of the initial Lorentz factor $\Gamma
_c$. The solid lines from the bottom to the top correspond to
$\Gamma _c=10,20,50,100,200,500,1000 $, respectively, and the
dotted lines from the bottom to the top are the $\Gamma =10$ and
$\Gamma =100$ lines presented in Fig. 1. Other parameters are the
same as those adopted in Fig. 12.} \label{Fig17}
\end{figure}

\begin{figure}[tbp]
\vbox to 3.0in{\rule{0pt}{3.0in}} \includegraphics{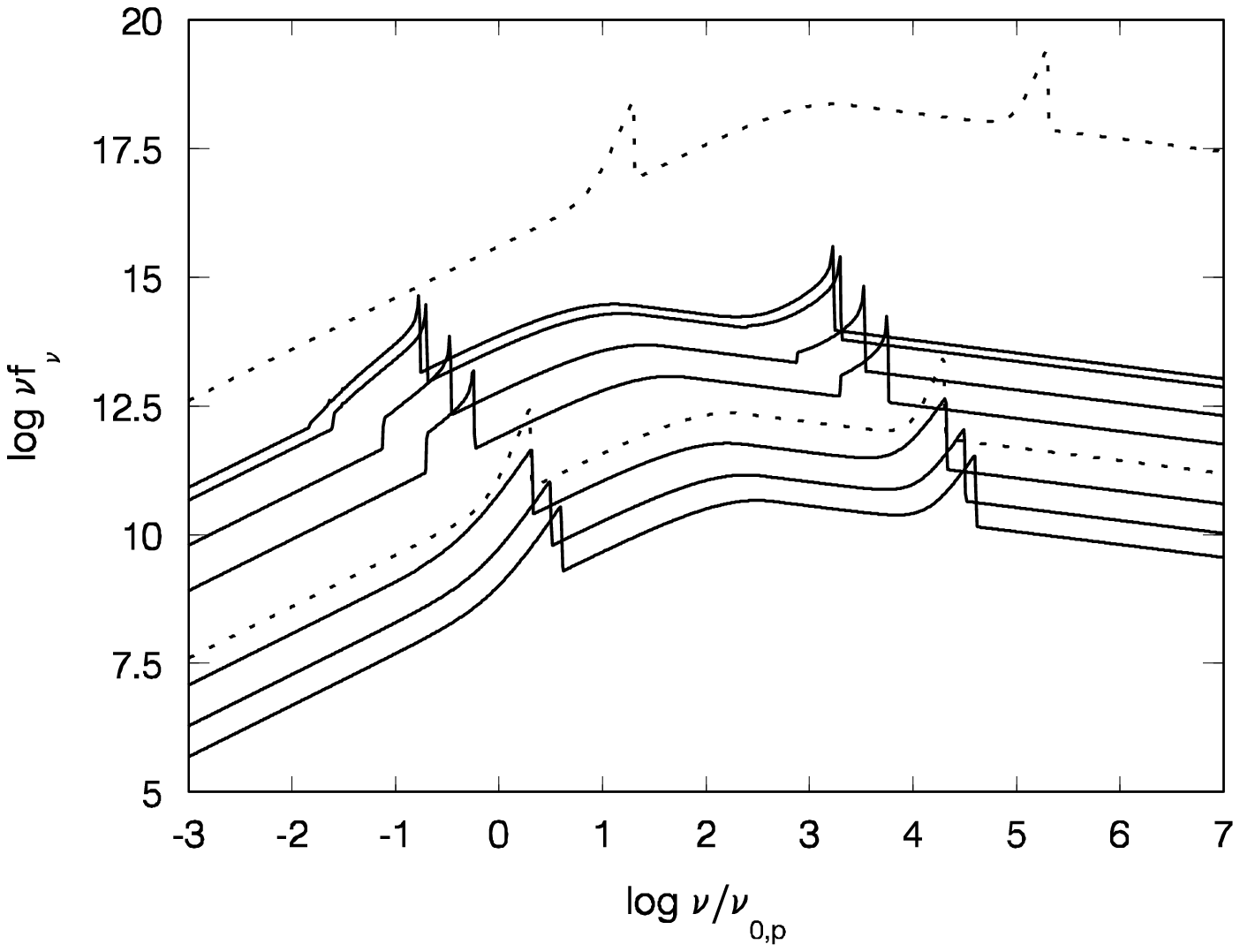} \caption{The
expected spectrum of a fireball with a variable Lorentz factor,
where we take $\Gamma _c=300$ and $k=10k_0(p=10,\Gamma
_c=300,R_c/c=1)$ and consider different observation times, which
are represented by different values of $p$, respectively. The
solid lines from the bottom to the top correspond to
$p=0,1,3,10,100,1000,2000$, respectively, and the dotted lines
from the bottom to the top are the $\Gamma =100$ and $\Gamma
=1000$ lines presented in Fig. 1. Other parameters are the same as
those adopted in Fig. 12.} \label{Fig18}
\end{figure}

We find in Fig. 14 an unfamiliar form of emission lines. To view
this phenomenon in more detail, see Fig. 17, where more lines with
different values of $\Gamma _c$ are presented (a detailed study of
this issue will be made later). It shows that, for the assigned
observation time, the emission lines concerned would be sharp on
both edges in some cases where the initial Lorentz factor is
carefully chosen. To find out if this phenomenon depends also on
the observation time, we plot Fig. 18, where different observation
times are considered for a single initial Lorentz factor. One
finds that the shape of the lines is indeed changed with time.
During some period, it would be sharp on both edges. Meanwhile,
the flux keeps being enhanced (at least during the epoch of the
deceleration phase concerned). This would become an interesting
expected observational characteristic.

\section{Discussion and Conclusions}

In this paper, we investigate how the Doppler effect in the
fireball framework affects the observed spectrum of the object
with various rest frame emission or absorption lines.

In section 2, we study the effect on narrow emission lines. The
study reveals that, influenced by the Doppler effect in the
fireball framework, rest frame narrow emission lines would
significantly shift to higher energy bands and would be
significantly broadened. A rest frame narrow emission line at high
energy bands could form an up-turning tip of the high energy tail
in the observed spectrum. It shows that, when $\Gamma $ is large
enough, the observed line frequency $\nu _{line}$ and the rest
frame line frequency $\nu _{0,line}$ would be related by $\nu
_{line}\approx 2\Gamma \nu _{0,line}$. It also reveals that the
relative width of the line feature would approach an asymptotic
value when $\Gamma \gg 1$. In the case where the relative width of
the rest frame narrow emission lines is $0.0235$, the observed
$\Delta \nu _{FWHM}/\nu _{line}$ would become almost one magnitude
larger than that of the rest frame emission line. When the rest
frame line is narrow enough, the relative width would approach
$0.162$.

In section 3, the effects on other lines are investigated. The
study shows that in the spectrum of a relativistically expanding
fireball, any rest frame lines would shift to higher energy bands
and would be significantly smoothed. Rest frame weak narrow
emission lines as well as narrow absorption lines and absorption
line forests would be smoothed and would hardly be detectable.
Meanwhile, the features of rest frame broad emission lines as well
as both strong and weak broad absorption lines would remain almost
unchanged and therefore would be easier to detect. Deep gaps
caused by rest frame broad absorption lines would be significantly
filled. In addition, each of the rest frame emission line forests
would form a single broad line feature.

In section 4, we investigate the effect of time dependence of the
line intensity and the effect of variation of the Lorentz factor.
The study reveals that, in the spectrum of a relativistically
expanding fireball, rest frame dimming narrow emission lines would
also shift to higher energy bands and would be significantly
broadened. However, when the Lorentz factor is very large (say,
$\Gamma >1000$), the line feature would disappear. As time
progresses, the feature would eventually disappear and this would
occur at a much earlier time for those fireballs with a larger
value of $\Gamma $. Considering the variation of the Lorentz
factor of fireballs we find that the flux observed at a relatively
later time would be larger than that observed at an earlier time,
suggesting that, in the situation studied here, the enhancement of
the radius of the fireball plays an important role in determining
the magnitude of the flux. When the initial Lorentz factor is very
small, the effect of deceleration would be insignificant, and the
positions of emission lines would remain almost unchanged. When
the initial Lorentz factor is large enough, due to the
deceleration, the flux would become much smaller than the constant
radiation flux and the change of the positions of emission lines
would be obvious. As time progresses, the positions would shift to
lower energy bands. In addition, we find an unfamiliar form of
emission lines that is sharp on both edges. Further study reveals
that this phenomenon depends not only on the initial Lorentz
factor but also on the observation time. For some initial Lorentz
factors, the shape of lines changes with time. This might become
an interesting observational characteristic expected in GRBs.

One can conclude that, for a relativistically expanding fireball,
there would be no narrow emission lines expected in its spectrum,
and rest frame narrow absorption lines as well as rest frame
emission or absorption line forests would not be detectable. Since
$\nu _{line}$ is a linear function of $\Gamma $, features of lines
could serve as an indicator of the expansion speed of fireballs as
long as they are identified. In the phase of deceleration of the
Lorentz factor, one can expect a change of the shape of emission
lines of fireballs with time.

If there are some high energy emission lines such as the $6.4keV$
line and the $ 511keV$ annihilation line in the outer shell of
fireballs, we could expect them to appear at around $ 0.1MeV$ and
$10MeV$ respectively when $\Gamma \sim 10$, or $1MeV$ and $100MeV
$ when $\Gamma \sim 100$, or $10MeV$ and $1000MeV$ when $\Gamma
\sim 1000$. Regarding the previous report of the detection of
emission features near $ 400keV$ (Mazets et al. 1980), we would
prefer to interpret them as the blue-shifted $6.4keV$ line
corresponding to the fireball expanding with $\Gamma \sim 30$. Due
to the great blue shift effect, the conventional redshift
interpretation of the line seems unlikely. We propose that, in
detecting line features, the blue shift and broadening effects due
to the expansion of the fireball should be taken into account.

In addition to the conventional fireball model, the method can
also be applied to the cannonball model (Dado et al. 2002a,
2002b). To explain the X-ray line features observed in some GRBs
as being Doppler-boosted, Dado et al. (2002c) employed a
cannonball model for which the radius of the cannonball was
treated as a constant and the direction of the motion of the
cannonball was assumed to be unchanged. In this situation, some
formulas in Paper I can be
simply applied. Taking $t_{0,\theta }-t_{0,c}$ as the proper emission time $%
t_0$ and $t-t_c-D/c+(R_c/c)\cos \theta $ as the observation time
$t$ we would get from (A.8) of Paper I that $t_0=\delta (t)t$,
where $\delta (t)\equiv 1/\Gamma (t)(1-\beta (t)\cos \theta )$ is
the Doppler factor of the cannonball at time $t$. The observed
area of the cannonball is $S(\theta ,\Gamma )R^2$, where $S(\theta
,\Gamma )$ is a function of the Lorentz factor, which would be
$\pi $ when the contraction effect due to the relative motion is
ignored. Replacing $R^2(t_\theta )\cos \theta \sin \theta d\theta
d\varphi $ with $S(\theta ,\Gamma )R^2$ we get from (A.11) of
Paper I that $dE=\Delta s_{ob}d\nu dtS(\theta ,\Gamma )R^2I_\nu
(\nu ,t)/D^2$, where the cosmological effect is ignored. Absorbing
$S(\theta ,\Gamma )$ into the normalized coefficient we get
$F_{obs}(\nu ,t)=R^2I_\nu (\nu ,t)/D^2 $. Applying the Doppler
effect and replacing $I_\nu (\nu ,t)$ with $\delta ^3(t)I_{0,\nu
}(\nu _0,t_0)=\delta ^3(t)F_{CB}[\nu /\delta (t),\delta (t)t]$ we
come to equation (3) of Dado et al. (2002c) and the Doppler effect
itself leads to equation (6) there (when the cosmological effect
is ignored); based on this the spectral data (including the X-ray
line features) of some GRBs were fitted (Dado et al. 2002c).

\vspace{8mm}

{\sl Acknowlegements}. It is my great pleasure to thank Prof. M.
J. Rees for his helpful suggestions and discussion. This work was
supported by the Special Funds for Major State Basic Research
Projects (``973'') and National Natural Science Foundation of
China (No. 10273019).

\appendix

\section{Analytical study of $\Delta \nu _{FWHM}/\nu _{line}$\ for very
narrow emission lines}

Here we calculate $\Delta \nu _{FWHM}/\nu _{line}$ analytically
for very narrow emission lines when $\Gamma \gg 1$.

The flux expected from a fireball with a definite value of the
Lorentz factor is (see Paper I)
\begin{equation}
f_\nu (t)=\frac{2\pi \widetilde{R}^2(t)}{D^2\Gamma ^3}\int_{\theta
_{\min }}^{\theta _{\max }}\frac{I_{0,\nu }(t_{0,\theta },\nu
_{0,\theta })\cos \theta \sin \theta }{(1-\beta \cos \theta
)^5}d\theta ,
\end{equation}
where $\widetilde{R}(t)$ is determined by (2); $t$, $\nu $, $D$, $\theta $, $%
\nu _{0,\theta }$, $t_{0,\theta }$, and $I_{0,\nu }(t_{0,\theta
},\nu _{0,\theta })$ are explained in section 2. As discussed in
section 2, we consider a constant radiation and get the following
by applying (3):
\begin{equation}
f_\nu (t)=\frac{2\pi I_0\widetilde{R}^2(t)}{D^2\Gamma
^3}\int_{\theta _{\min }}^{\theta _{\max }}g_{0,\nu }(\nu
_{0,\theta })\frac{\cos \theta \sin \theta }{(1-\beta \cos \theta
)^5}d\theta .
\end{equation}

To make an analytical study of $\Delta \nu _{FWHM}/\nu _{line}$
for very narrow emission lines, one would prefer a simple form of
emission, which can be calculated analytically and can represent
an emission line in extreme cases.

Let us consider a very narrow rectangle of emission. The rest
frame radiation is assumed to be
\begin{equation}
g_{0,\nu }(\nu _{0,\theta })=g_{0,\nu ,c}(\nu _{0,\theta
})+e_{0,\nu }(\nu _{0,\theta }),
\end{equation}
where, while $g_{0,\nu ,c}(\nu _{0,\theta })$ describes the
relative intensity of a continuum (see section 2), $e_{0,\nu }(\nu
_{0,\theta })$ stands for the relative intensity of an emission
line which is
\begin{equation}
e_{0,\nu }(\nu _{0,\theta })=h\qquad \qquad \qquad \qquad (\nu
_{0,line}\leq \nu _{0,\theta }\leq \nu _{0,line}+\triangle \nu
_{0,line}),
\end{equation}
where $h$ is a constant.

To calculate $\Delta \nu _{FWHM}/\nu _{line}$, one needs only to
deal with the radiation of the line:
\begin{equation}
f_{\nu ,line}(t)=\frac{2\pi I_0\widetilde{R}^2(t)}{D^2\Gamma
^3}e_\nu
\end{equation}
with
\begin{equation}
e_\nu \equiv h\int_{\theta _{\min }}^{\theta _{\max }}\frac{\cos
\theta \sin \theta }{(1-\beta \cos \theta )^5}d\theta .
\end{equation}

According to the Doppler effect, for certain values of $\nu
_{0,\theta }$ and $\Gamma $, the shifted frequency $\nu $ is
confined in the range $\nu _{0,\theta }/\Gamma \leq \nu \leq \nu
_{0,\theta }/\Gamma (1-\beta )$ for an expanding fireball. Thus,
the valid range of frequency in (A.6) is
\begin{equation}
\frac{\nu _{0,line}}\Gamma \leq \nu \leq \frac{\nu
_{0,line}+\Delta \nu _{0,line}}{\Gamma (1-\beta )}.
\end{equation}
The integral limits $\theta _{\min }$ and $\theta _{\max }$ in
(A.6) can be determined by (Paper I)
\begin{equation}
\begin{tabular}{c}
$\theta _{\min }=\cos ^{-1}\left( \min \{1,\frac 1\beta
(1-\frac{\nu
_{0,line}}{\Gamma \nu })\}\right) $ \\
$\qquad \qquad \qquad \qquad \qquad \left( 0\leq \frac 1\beta
(1-\frac{\nu
_{0,line}}{\Gamma \nu });0<\beta \right) $%
\end{tabular}
\end{equation}
and
\begin{equation}
\begin{tabular}{c}
$\theta _{\max }=\cos ^{-1}\left( \max \{0,\frac 1\beta
(1-\frac{\nu
_{0,line}+\triangle \nu _{0,line}}{\Gamma \nu })\}\right) $ \\
$\qquad \qquad \qquad \qquad \qquad \left( \frac 1\beta
(1-\frac{\nu
_{0,line}+\triangle \nu _{0,line}}{\Gamma \nu })\leq 1;0<\beta \right) .$%
\end{tabular}
\end{equation}

Integrating (A.6) we get
\begin{equation}
\begin{array}{c}
e_\nu =\frac h{\beta ^2}\{\frac 14[\frac 1{(1-\beta \min \{1,\frac 1\beta (1-%
\frac{\nu _{0,line}}{\Gamma \nu })\})^4}-\frac 1{(1-\beta \max
\{0,\frac 1\beta (1-\frac{\nu _{0,line}+\triangle \nu
_{0,line}}{\Gamma \nu })\})^4}]
\\
+\frac 13[\frac 1{(1-\beta \max \{0,\frac 1\beta (1-\frac{\nu
_{0,line}+\triangle \nu _{0,line}}{\Gamma \nu })\})^3}-\frac
1{(1-\beta \min
\{1,\frac 1\beta (1-\frac{\nu _{0,line}}{\Gamma \nu })\})^3}]\} \\
\qquad \qquad \qquad \qquad \qquad \qquad \qquad \qquad \qquad
(\frac{\nu
_{0,line}}\Gamma \leq \nu \leq \frac{\nu _{0,line}+\Delta \nu _{0,line}}{%
\Gamma (1-\beta )}).
\end{array}
\end{equation}

Let us calculate $\partial e_\nu /\partial \nu $ in different
frequency ranges: range I, $(\frac{\nu _{0,line}}\Gamma \leq \nu
\leq \frac{\nu _{0,line}+\Delta \nu _{0,line}}\Gamma )$; range II,
$(\frac{\nu
_{0,line}+\triangle \nu _{0,line}}\Gamma \leq \nu \leq \frac{\nu _{0,line}}{%
\Gamma (1-\beta )})$; and range III, $(\frac{\nu _{0,line}}{\Gamma (1-\beta )}%
\leq \nu \leq \frac{\nu _{0,line}+\Delta \nu _{0,line}}{\Gamma (1-\beta )})$%
. Here, we assume
\begin{equation}
\frac{\triangle \nu _{0,line}}{\nu _{0,line}}\ll 1\qquad \qquad
and\qquad \qquad 1\ll \Gamma ,
\end{equation}
which leads to
\begin{equation}
\nu _{0,line}+\triangle \nu _{0,line}\ll \frac{\nu
_{0,line}}{1-\beta }.
\end{equation}
Applying (A.12), we find $\partial e_\nu /\partial \nu \geq 0$ in
both ranges I and II, and $\partial e_\nu /\partial \nu <0$ in
range III. This leads to
\begin{equation}
e_{\nu ,\max }=\frac h{3\beta ^2(1-\beta )^3}\{\frac{3[1-\frac{\nu
_{0,line}^4}{(\nu _{0,line}+\triangle \nu _{0,line})^4}]}{4(1-\beta )}+\frac{%
\nu _{0,line}^3}{(\nu _{0,line}+\triangle \nu _{0,line})^3}-1\},
\end{equation}
which is the maximum value of $e_\nu $ in the whole frequency
range.

The frequency at which $e_{\nu ,\max }$ is found is identified as
the observed line frequency $\nu _{line}$ which is determined by
\begin{equation}
\nu _{line}=\frac{\nu _{0,line}}{\Gamma (1-\beta )}.
\end{equation}
When $\Gamma \gg 1$, it would approach $\nu _{line}\simeq 2\Gamma
\nu _{0,line}$.

With (A.13), we first calculate $\nu _{FWHM}$ in region II, $\nu _{FWHM,II}$%
, and obtain
\begin{equation}
\begin{array}{c}
\frac{\Gamma ^4\nu _{FWHM,II}^4}{4\nu _{0,line}^4}[1-\frac{\nu _{0,line}^4}{%
(\nu _{0,line}+\triangle \nu _{0,line})^4}]-\frac 1{8(1-\beta )^4}[1-\frac{%
\nu _{0,line}^4}{(\nu _{0,line}+\triangle \nu _{0,line})^4}] \\
=\frac{\Gamma ^3\nu _{FWHM,II}^3}{3\nu _{0,line}^3}[1-\frac{\nu _{0,line}^3}{%
(\nu _{0,line}+\triangle \nu _{0,line})^3}]-\frac 1{6(1-\beta )^3}[1-\frac{%
\nu _{0,line}^3}{(\nu _{0,line}+\triangle \nu _{0,line})^3}].
\end{array}
\end{equation}
The solution of (A.15) can be approximated by
\begin{equation}
\frac{\Gamma \nu _{FWHM,II}}{\nu _{0,line}}\simeq \frac{%
[1+3(2^{1/12}-1)]^{1/3}}{2^{1/3}(1-\beta )}[1+\frac{2(2^{1/12}-1)}{%
1+3(2^{1/12}-1)}\frac{\triangle \nu _{0,line}}{\nu _{0,line}}].
\end{equation}
Second, we calculate $\nu _{FWHM}$ in region III, $\nu
_{FWHM,III}$, and get
\begin{equation}
\begin{array}{c}
\frac{\Gamma ^4\nu _{FWHM,III}^4}{4\nu _{0,line}^4}-\frac{1+(1+\frac{%
\triangle \nu _{0,line}}{\nu _{0,line}})^4}{8(1-\beta )^4} \\
=\frac{\Gamma ^3\nu _{FWHM,III}^3(1+\frac{\triangle \nu
_{0,line}}{\nu _{0,line}})}{3\nu
_{0,line}^3}-\frac{(1+\frac{\triangle \nu _{0,line}}{\nu
_{0,line}})[1+(1+\frac{\triangle \nu _{0,line}}{\nu _{0,line}})^3]}{%
6(1-\beta )^3}.
\end{array}
\end{equation}
The solution of (A.17) can be approximated by
\begin{equation}
\frac{\Gamma \nu _{FWHM,III}}{\nu _{0,line}}\simeq \frac
1{2^{1/3}(1-\beta
)}(2^{1/3}+2^{-2/3}\frac{\triangle \nu _{0,line}}{\nu _{0,line}})=\frac{%
1+\frac 12\frac{\triangle \nu _{0,line}}{\nu _{0,line}}}{1-\beta
}.
\end{equation}
We therefore obtain
\begin{equation}
\begin{array}{c}
\frac{\Delta \nu _{FWHM}}{\nu _{line}}\simeq 1-[\frac{1+3(2^{1/12}-1)}%
2]^{1/3}+\{\frac 12-\frac{2^{2/3}(2^{1/12}-1)}{[1+3(2^{1/12}-1)]^{2/3}}\}%
\frac{\triangle \nu _{0,line}}{\nu _{0,line}} \\
\simeq 0.162+0.415\frac{\triangle \nu _{0,line}}{\nu _{0,line}}.
\end{array}
\end{equation}
When $\triangle \nu _{0,line}/\nu _{0,line}\rightarrow 0$, one will get $%
\Delta \nu _{FWHM}/\nu _{line}\simeq 0.162$.

\section{Expected flux of a fireball with a variable Lorentz factor}

We examine a fireball expanding at a variable velocity $v=\beta c$
and employ the same coordinate system adopted in section 2.
Consider radiation from a rest frame differential surface,
$ds_{0,\theta ,\varphi }$, of the fireball at proper time
$t_{0,\theta }$. Let $ds_{\theta ,\varphi }$ be the corresponding
differential surface resting on the observer's framework,
coinciding with $ds_{0,\theta ,\varphi }$ at $t_{0,\theta }$, and
let $ t_\theta $ be the corresponding coordinate time of
$t_{0,\theta }$. Obviously, $t_\theta =t_\theta (t_{0,\theta })$.

According to the theory of special relativity, $t_\theta $ and
$t_{0,\theta } $ are related by
\begin{equation}
dt_\theta =\Gamma (t_\theta )dt_{0,\theta }=\Gamma _0(t_{0,\theta
})dt_{0,\theta },
\end{equation}
where $\Gamma (t_\theta )$ is the Lorentz factor of the fireball at $%
t_\theta $ and $\Gamma _0(t_{0,\theta })\equiv \Gamma [t_\theta
(t_{0,\theta })]$. Integrating (B.1) yields
\begin{equation}
t_\theta =\int_{t_{0,c}}^{t_{0,\theta }}\Gamma _0(t_{0,\theta
})dt_{0,\theta }+t_c\qquad \qquad or\qquad \qquad t_{0,\theta
}=\int_{t_c}^{t_\theta }\frac 1{\Gamma (t_\theta )}dt_\theta
+t_{0,c},
\end{equation}
here we assign $t_\theta =$ $t_c$ when $t_{0,\theta }=t_{0,c}$.

The area of $ds_{\theta ,\varphi }$ is
\begin{equation}
ds_{\theta ,\varphi }=R^2(t_\theta )\sin \theta d\theta d\varphi ,
\end{equation}
where $R(t_\theta )$ is the radius of the fireball at $t_\theta $.
Suppose the radius develops as
\begin{equation}
dR(t_\theta )=c\beta (t_\theta )dt_\theta .
\end{equation}
Integrating (B.4) leads to
\begin{equation}
R(t_\theta )=c\int_{t_c}^{t_\theta }\beta (t_\theta )dt_\theta
+R_c,
\end{equation}
where $R_c$ is the radius at time $t_\theta =t_c$. As assigned
above, $ t_\theta $ and $t_{0,\theta }$ correspond to the same
moment. Therefore the radius can be expressed as
\begin{equation}
R_0(t_{0,\theta })\equiv R[t_\theta (t_{0,\theta
})]=c\int_{t_c}^{t_\theta (t_{0,\theta })}\beta (t_\theta
)dt_\theta +R_c=c\int_{t_{0,c}}^{t_{0,\theta }}\beta
_0(t_{0,\theta })\Gamma _0(t_{0,\theta })dt_{0,\theta }+R_c,
\end{equation}
where (B.1) is applied, and $\beta _0(t_{0,\theta })\equiv \beta
[t_\theta (t_{0,\theta })]$.

Let us consider an observation within small intervals $t$ --- $t+dt$ and $%
\nu $ --- $\nu +d\nu $ carried out by an observer with a detector
$\triangle s_{ob}$ at a distance $D$ ($D$ is the distance between
the observer and the
center of the fireball), where $D\gg R(t_\theta )$. Suppose radiation from $%
ds_{0,\theta ,\varphi }$ reaching the observer within the above
observation
intervals is emitted within the proper time interval $t_{0,\theta }$ --- $%
t_{0,\theta }+dt_{0,\theta }$ and the rest frame frequency
interval $\nu _{0,\theta }$ --- $\nu _{0,\theta }+d\nu _{0,\theta
}$. According to the Doppler effect, $\nu $ and $\nu _{0,\theta }$
are related by
\begin{equation}
\nu =\frac{\nu _{0,\theta }}{\Gamma (t_\theta )[1-\beta (t_\theta
)\cos \theta ]}=\frac{\nu _{0,\theta }}{\Gamma _0(t_{0,\theta
})[1-\beta _0(t_{0,\theta })\cos \theta ]}.
\end{equation}

Considering the travelling of light from the fireball to the
observer we get
\begin{equation}
c(t-t_\theta )=D-R(t_\theta )\cos \theta
\end{equation}
(here the cosmological effect is ignored). Combining (B.5) and
(B.8) we get
\begin{equation}
\lbrack \int_{t_c}^{t_\theta }\beta (t_\theta )dt_\theta ]\cos
\theta -t_\theta =D/c-t-(R_c/c)\cos \theta ,
\end{equation}
while combining (B.2), (B.6) and (B.8) we obtain
\begin{equation}
\lbrack \int_{t_{0,c}}^{t_{0,\theta }}\beta _0(t_{0,\theta
})\Gamma _0(t_{0,\theta })dt_{0,\theta }]\cos \theta
-[\int_{t_{0,c}}^{t_{0,\theta }}\Gamma _0(t_{0,\theta
})dt_{0,\theta }]=D/c-t-(R_c/c)\cos \theta +t_c.
\end{equation}
Once $\beta (t_\theta )$ or $\beta _0(t_{0,\theta })$ is known,
$t_\theta $ and $t_{0,\theta }$ as functions of $\theta $ and $t$
can be determined by (B.9) and (B.10), respectively. Thus the
radius of the fireball can be determined by (B.5) or (B.6).

Suppose photons, which are emitted from $ds_{0,\theta ,\varphi }$
within the proper time interval $t_{0,\theta }$ --- $t_{0,\theta
}+dt_{0,\theta }$ and then reach the observer within $t$ ---
$t+dt$, pass through $ds_{\theta ,\varphi }$ within the coordinate
time interval $t_\theta $ --- $t_\theta +dt_{\theta ,s}$. Since
both $ds_{\theta ,\varphi }$ and the observer rest in the same
framework, $dt_{\theta ,s}=dt$. Of course, the frequency interval
for the photons measured by both $ds_{\theta ,\varphi
}$ and the observer must be the same, which is taken as $\nu $ --- $\nu $ $%
+d\nu $. In views of $ds_{\theta ,\varphi }$, the amount of energy
emitted from $ds_{0,\theta ,\varphi }$ within $t_{0,\theta }$ ---
$t_{0,\theta }+dt_{0,\theta }$ and $\nu _{0,\theta }$ --- $\nu
_{0,\theta }+d\nu _{0,\theta }$ towards the observer would be
\begin{equation}
dE_{\theta ,\varphi }=I_\nu (t_\theta ,\nu ,\theta )\cos \theta
ds_{\theta ,\varphi }d\omega d\nu dt,
\end{equation}
where $I_\nu (t_\theta ,\nu ,\theta )$ is the intensity of
radiation measured by $ds_{\theta ,\varphi }$, which is allowed to
be a function of direction, and $d\omega $ is the solid angle of
$\triangle s_{ob}$ with respect to the fireball, which is
\begin{equation}
d\omega =\frac{\triangle s_{ob}}{D^2}.
\end{equation}
Thus,
\begin{equation}
dE_{\theta ,\varphi }=\frac{\triangle s_{ob}d\nu dtR^2(t_\theta
)I_\nu (t_\theta ,\nu ,\theta )\cos \theta \sin \theta d\theta
d\varphi }{D^2},
\end{equation}
where (B.3) and (B.12) are applied.

The total amount of energy emitted from the whole fireball surface
detected
by the observer within the above observation intervals is an integral of $%
dE_{\theta ,\varphi }$ over that area, which is
\begin{equation}
dE=\frac{2\pi \triangle s_{ob}d\nu dt}{D^2}\int_{\theta _{\min
}}^{\theta _{\max }}R^2(t_\theta )I_\nu (t_\theta ,\nu ,\theta
)\cos \theta \sin \theta d\theta ,
\end{equation}
where $\theta _{\min }$ and $\theta _{\max }$ are determined by
the fireball surface itself together with the emitted ranges of
$t_{0,\theta }$ and $\nu _{0,\theta }$. Thus, the expected flux
would be
\begin{equation}
f_\nu (t)=\frac{2\pi }{D^2}\int_{\theta _{\min }}^{\theta _{\max
}}R^2(t_\theta )I_\nu (t_\theta ,\nu ,\theta )\cos \theta \sin
\theta d\theta .
\end{equation}

It is well known that the observer frame intensity $I_\nu
(t_\theta ,\nu ,\theta )$ is related to the rest frame intensity
$I_{0,\nu }(t_{0,\theta },\nu _{0,\theta },\theta )$ by
\begin{equation}
I_\nu (t_\theta ,\nu ,\theta )=(\frac \nu {\nu _{0,\theta
}})^3I_{0,\nu }(t_{0,\theta },\nu _{0,\theta },\theta ).
\end{equation}
The flux then can be written as
\begin{equation}
f_\nu (t)=\frac{2\pi }{D^2}\int_{\theta _{\min }}^{\theta _{\max }}\frac{%
R_0^2(t_{0,\theta })I_{0,\nu }(t_{0,\theta },\nu _{0,\theta
},\theta )\cos \theta \sin \theta }{\Gamma _0^3(t_{0,\theta
})[1-\beta _0(t_{0,\theta })\cos \theta ]^3}d\theta ,
\end{equation}
where (B.6) and (B.7) are applied.

The range of $\theta $ of the visible fireball surface is
\begin{equation}
0\leq \theta \leq \pi /2.
\end{equation}
Within this range, suppose the emitted ranges of $t_{0,\theta }$
and $\nu _{0,\theta }$ constrain $\theta $ by
\begin{equation}
\theta _{t,\min }\leq \theta \leq \theta _{t,\max }
\end{equation}
and
\begin{equation}
\theta _{\nu ,\min }\leq \theta \leq \theta _{\nu ,\max },
\end{equation}
respectively. Then when the following condition
\begin{equation}
\max \{\theta _{t,\min },\theta _{\nu ,\min }\}<\min \{\theta
_{t,\max },\theta _{\nu ,\max }\}
\end{equation}
is satisfied, $\theta _{\min }$ and $\theta _{\max }$ would be
obtained by
\begin{equation}
\theta _{\min }=\max \{\theta _{t,\min },\theta _{\nu ,\min }\}
\end{equation}
and
\begin{equation}
\theta _{\max }=\min \{\theta _{t,\max },\theta _{\nu ,\max }\},
\end{equation}
respectively.

Let the emitted ranges of $t_{0,\theta }$ and $\nu _{0,\theta }$
be
\begin{equation}
t_{0,\min }\leq t_{0,\theta }\leq t_{0,\max }
\end{equation}
and
\begin{equation}
\nu _{0,\min }\leq \nu _{0,\theta }\leq \nu _{0,\max },
\end{equation}
respectively. For $\beta >0$, one can obtain the following from
(B.7) and (B.25):
\begin{equation}
\begin{tabular}{c}
$\theta _{\nu ,\min }(t_{0,\theta })=\cos ^{-1}\left( \min
\{1,\frac 1\beta
(1-\frac{\nu _{0,\min }}{\Gamma \nu })\}\right) $ \\
$\qquad \qquad \qquad \qquad \qquad \left( 0\leq \frac 1\beta
(1-\frac{\nu
_{0,\min }}{\Gamma \nu });0<\beta \right) $%
\end{tabular}
\end{equation}
and
\begin{equation}
\begin{tabular}{c}
$\theta _{\nu ,\max }(t_{0,\theta })=\cos ^{-1}\left( \max
\{0,\frac 1\beta
(1-\frac{\nu _{0,\max }}{\Gamma \nu })\}\right) $ \\
$\qquad \qquad \qquad \qquad \qquad \left( \frac 1\beta
(1-\frac{\nu
_{0,\max }}{\Gamma \nu })\leq 1;0<\beta \right) ,$%
\end{tabular}
\end{equation}
where $\beta =\beta (t_\theta )=\beta _0(t_{0,\theta })$ and
$\Gamma =\Gamma (t_\theta )=\Gamma _0(t_{0,\theta })$. From (B.10)
and (B.24) one can also determine $\theta _{t,\min }$ and $\theta
_{t,\max }$ once $\beta (t_\theta ) $ or $\beta _0(t_{0,\theta })$
is known.

\section{Lower limit of the coefficient of deceleration}

We consider in section 4 the case of the radiative hydrodynamics
for which, when decelerated, the development of the Lorentz factor
of a fireball follows that shown by (44). The coordinate time,
$t_\theta $, of the
differential surface of the fireball concerned and the observation time, $t$%
, are related by (B.9). Since $\beta (t_\theta )\leq \beta _c$,
where $\beta _c\equiv \sqrt{\Gamma _c^2-1}/\Gamma _c$, we obtain
from (B.9) that
\begin{equation}
t_\theta \leq \frac{t-\frac Dc+(\frac{R_c}c-t_c\beta _c)\cos \theta }{%
1-\beta _c\cos \theta }.
\end{equation}
Applying (49) we get
\begin{equation}
t_\theta \leq t_c+\frac{R_c}c\frac{p+\cos \theta }{1-\beta _c\cos
\theta }.
\end{equation}

To focus on the effect of variation of the Lorentz factor, we
require that all photons observed at the assigned time must be
those emitted after the deceleration starts and before the
expansion stops, i.e.,
\begin{equation}
t_\theta \geq t_c
\end{equation}
and
\begin{equation}
\Gamma (t_\theta )>1.
\end{equation}

Combining (C.2) and (C.3) we get
\begin{equation}
p\geq 0.
\end{equation}

One can verify by applying (44) and (45) that, when
\begin{equation}
k^{7/3}>t_c+\frac{R_c}c\frac{p+\cos \theta }{1-\beta _c\cos \theta
},
\end{equation}
then (C.4) would be satisfied. Taking $\theta =0$ we get
\begin{equation}
k^{7/3}>t_c+\frac{R_c}c\frac{1+p}{1-\beta _c}.
\end{equation}
Applying (45) we arrive at
\begin{equation}
k>\Gamma _c[\frac{R_c(1+p)}{c(1-\beta _c)(\Gamma
_c^{7/3}-1)}]^{3/7}.
\end{equation}

\clearpage

\end{document}